\documentclass[longbibliography,prb,aps,twocolumn,showpacs,floatfix]{revtex4-2}
\usepackage{listings}
\usepackage{amsfonts}
\usepackage{amssymb}
\usepackage{hyperref}
\usepackage{graphicx}
\usepackage{dcolumn}
\usepackage{bm,amsmath,verbatim}
\usepackage{mathrsfs}
\usepackage{color}
\usepackage{lineno}
\usepackage{dsfont}
\usepackage{amsmath,amssymb,amsthm}
\usepackage{float}

\usepackage[T1]{fontenc}
\usepackage[latin9]{inputenc}
\usepackage{booktabs}

\renewcommand\toprule{\hline\hline}
\renewcommand\bottomrule{\hline\hline}

\hypersetup{colorlinks,
	linkcolor=blue,          citecolor=blue,        filecolor=blue,      urlcolor=blue           }
\def\be{\begin{equation}}
	\def\ee{\end{equation}}
\def\bea{\begin{eqnarray}}
	\def\eea{\end{eqnarray}}
\def\bse{\begin{subequations}}
	\def\ese{\end{subequations}}

\def\be{\begin{eqnarray}}
	\def\ee{\end{eqnarray}}

\begin{document}
	
%\setlength\columnsep{25pt}
%\linenumbers

\title{Observation of topological phases without crystalline counterparts}

\author{Mou Yan$^{1}$}
\thanks{These authors contributed equally to this work.}

\author{Yu-Liang Tao$^{2,1}$}
\thanks{These authors contributed equally to this work.}

\author{Yichong Hu$^{3}$}

\author{Zhenxing Cui$^{3}$}

\author{Jiong-Hao Wang$^{2,4}$}

\author{Gang Chen$^{1}$}
\email{chengang971@163.com}

\author{Yong Xu$^{2}$}
\email{yongxuphy@tsinghua.edu.cn}

\affiliation{$^{1}$Quantum Information Institute, School of Physics and Laboratory of Zhongyuan Light, Zhengzhou University, Zhengzhou 450001, China}

\affiliation{$^{2}$Center for Quantum Information, IIIS, Tsinghua University, Beijing 100084, China}

\affiliation{$^{3}$State Key Laboratory of Quantum Optics Technologies and Devices, Institute of Laser spectroscopy, Shanxi University, Taiyuan 030006, China}

\affiliation{$^{4}$Department of Physics, Stockholm University, AlbaNova University Center, 106 91 Stockholm, Sweden}

\begin{abstract}
Topological phases have been extensively studied primarily in crystalline 
systems with translational symmetry. Recent theoretical studies, however, have demonstrated
the existence of topological phases in quasicrystals that are absent in crystals. 
Here, we design a practically realizable tight-binding model on the Ammann-Beenker quasicrystalline lattice. 
This model respects eight-fold rotational and chiral symmetries, 
resulting in a higher-order topological phase with eight zero-energy corner modes that have no crystalline counterparts. 
We experimentally explore the topological phase in an acoustic quasicrystal. 
Surprisingly, we also discover symmetry-protected zero-energy modes near the center of the quasicrystal in a topologically trivial phase, 
a phenomenon not seen in crystals. We further experimentally observe these modes in a topologically trivial acoustic quasicrystal. 
\end{abstract}

\maketitle

\section{Introduction}
The discovery of quasicrystals in materials has revolutionized the field of crystallography
~\cite{shechtman1984metallic,levine1984quasicrystals,dubost1986large,steurer2004twenty}.
Unlike crystals with translational symmetry, quasicrystals are long-range ordered but not periodic.
This groundbreaking finding has sparked significant and enduring interest in the study of quasicrystals. 
Recent experimental advancements in this context include observation of quantum critical magnetic phenomena~\cite{deguchi2012quantum}, 
superconductivity~\cite{kamiya2018discovery,uri2023superconductivity,tokumoto2024superconductivity}, 
four-dimensional topology~\cite{lohse2018exploring,zilberberg2018photonic,tsesses2025four}, 
localization-delocalization transitions~\cite{wang2020localization,han2025observation}, and the Bose glass~\cite{yu2024observing}. 
One intriguing property of quasicrystals is the existence of new rotational symmetries,
including five, eight, ten, and twelve-fold ones~\cite{steurer2018quasicrystals,zoorob2000complete}, 
blessed by the  relaxation of translational symmetry.
This expansion enriches the rotational symmetries present in crystals, which are limited to two-fold, 
three-fold, four-fold and six-fold. 
From a more mathematical perspective, one can use a set of tiles to tesselate a plane without the requirement of translational symmetry, 
such as the Penrose tilling~\cite{Penrose1974Aesthetics} and Ammann-Beenker (AB) tiling~\cite{Beenker1982,gruenbaum1987},
which lead to qusicrystalline structures respecting five-fold and eight-fold rotational 
symmetries, respectively.

Topological phases of matter are an important class of states that extend beyond the Landau paradigm of symmetry breaking
~\cite{hasan2010colloquium,qi2011topological,chiu2016classification}. 
The intersection of topology and quasicrystals has led to various intriguing topological phenomena
~\cite{Zilberberg2012PRL,Chen2012PRL,Silberberg2013PRL,Kraus2013PRL,Bandres2016PRX,Loring2016PRL,Zilberberg2020Higher,Fulga2019PRL,Xu2020PRL,Cooper2020PRR,DHXu2020PRB,Liu2021NL,Huang2022FP,Huang2022PRL,Donghui2023PRB,mao2024higher,yang2024higher}. 
In particular, theoretical predictions indicate that the novel rotational symmetries present in quasicrystals can give 
rise to new higher-order topological phases absent in crystals~\cite{Fulga2019PRL,Xu2020PRL,Cooper2020PRR,DHXu2020PRB,Liu2021NL,Huang2022FP,Huang2022PRL,Donghui2023PRB,mao2024higher}. 
For example, a higher-order topological model constructed on the AB tiling quasicrystalline lattices
can host eight zero-energy corner modes, which is impossible in a crystal lattice~\cite{Fulga2019PRL,Xu2020PRL} 
(see Appendix A for details of the tight-binding model with its complicated hopping network). 
Metamaterials, such as acoustic lattices~\cite{han2025observation,susstrunk2015observation,lu2017observation,li2018weyl,xue2019acoustic,
Ding2019PRL,ni2020demonstration,Qi2020PRL,wei2021higher,luo2021observation,hu2021non,xi2025soft}, 
photonic lattices~\cite{freedman2006wave,wang2020localization,wang2009observation,zilberberg2018photonic,noh2018topological,yang2019realization,li2023exceptional,liu2025photonic} 
and electric circuits~\cite{lee2018topolectrical,imhof2018topolectrical,yang2024circuit}, 
have proven to be powerful tools for realizing topological phases~\cite{TP_RMP,xue2022topological,Bachtold20222RMP,ni2023topological,zhu2023topological,Chan2025PRL} due to their high controllability. For instance, 
the quadrupole insulator realized on an AB tiling quasicrystalline lattice with electric circuits~\cite{lv2021realization} 
possesses four-fold rotational symmetry and hosts four corner modes, similar to the Benalcazar-Bernevig-Hughes 
model on a square lattice~\cite{BBH1,BBH2}. Another experiment has 
successfully created a photonic Stampfli-type quasicrystal hosting topological corner modes protected by six-fold rotational symmetry~\cite{shi2024observation}, 
which can also occur in crystalline systems. 
Recently, an AB tiling quasicrystal photonic system has been experimentally realized, exhibiting both in-gap edge states and eight corner states~\cite{Yang2025APL}.

\begin{figure*}[ht]
	\includegraphics[width=\linewidth]{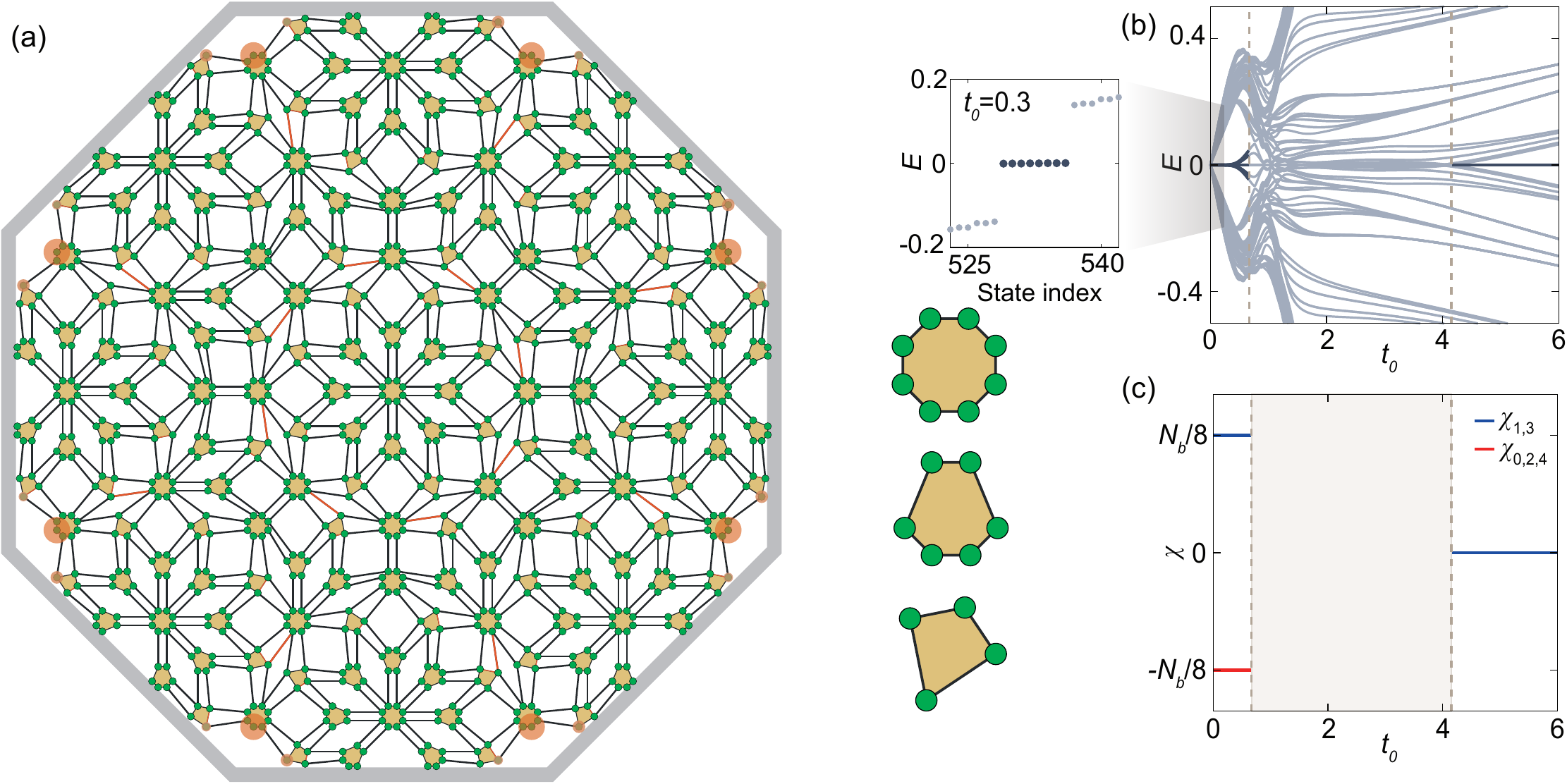}
	\caption{{Quasicrystalline structure gives rise to topological phases without crystalline counterparts.} 
		(a) Illustration of the tight-binding model in Eq.~\ref{H_TB} on an octagonal quasicrystal. 
		Each vertex of the quasicrystal hosts a cell, depicted as brown-filled polygons, 
		containing either eight, six, or four lattice sites.
		Connections between neighboring sites indicate the presence of hopping between these sites. 
		The bonds shown as red lines denote sign-reversal hoppings.
		The red-filled circles near the corners represent the spatial distribution of the local DOS 
		at zero energy, with the relative size of the circles indicating its magnitude.
		(b) Energy spectra of the tight-binding Hamiltonian in Eq.~\ref{H_TB} 
		on the quasicrystal with respect to the intra-cell hopping strength $t_0$. 
		The system exhibits a topological regime with eight corner modes for $0<t_0<0.67$ and a 
		trivial regime with two zero-energy modes for $t_0>4.15$.
		Between these regions, the bulk spectrum becomes gapless.
		Inset: the energy spectrum at $t_0=0.3$ with respect to the state index, 
		highlighting the presence of eight zero-energy states.
		(c) Topological invariants $\chi_p$ as a function of $t_0$ with $p=0,\dots,4$. 
		In the topological regime, $\chi_p=\pm N_b/8$, where $N_b$ is the number of 
		boundary sites. For the quasicrystalline structure shown in (a), $N_b=40$. 
	} 
	\label{Fig1}
\end{figure*}

Here we theoretically propose and experimentally realize a tight-binding model on an AB tiling quasicrystalline 
lattice. 
The model respects eight-fold rotational symmetry, a feature absent in crystals, as well as chiral symmetry. 
In contrast to previous models on the AB tiling quasicrystal~\cite{Fulga2019PRL,Xu2020PRL}, 
our model does not possess a generalized 
quadrupole moment~\cite{mao2024higher}. 
It also differs considerably from the model in Ref.~\cite{Yang2025APL}, whose topological properties remain unknown.
Interestingly, we find that our model can still be topological, 
with eight zero-energy corner modes lacking crystalline counterparts. 
To characterize this topological behavior, we introduce a rotational-symmetry-based invariant. 
In the experiment, we realize the model on an acoustic quasicrystal
with the quasicrystalline nature validated by Fourier spectrum of the measured acoustic pressure field. 
Meanwhile, we experimentally observe the presence of eight zero-energy corner modes.
Remarkably, we also observe two zero-energy modes localized 
at the center of the quasicrystal in a topologically trivial phase, 
which are protected by eight-fold rotational, time-reversal, and chiral symmetries.
To show the generality of our approach, 
we propose a decagonal quasicrystalline structure, which supports ten corner modes.

\section{Tight-binding model on the octagonal quasicrystalline structure}

\begin{figure*}[tp]
	\includegraphics[width=\linewidth]{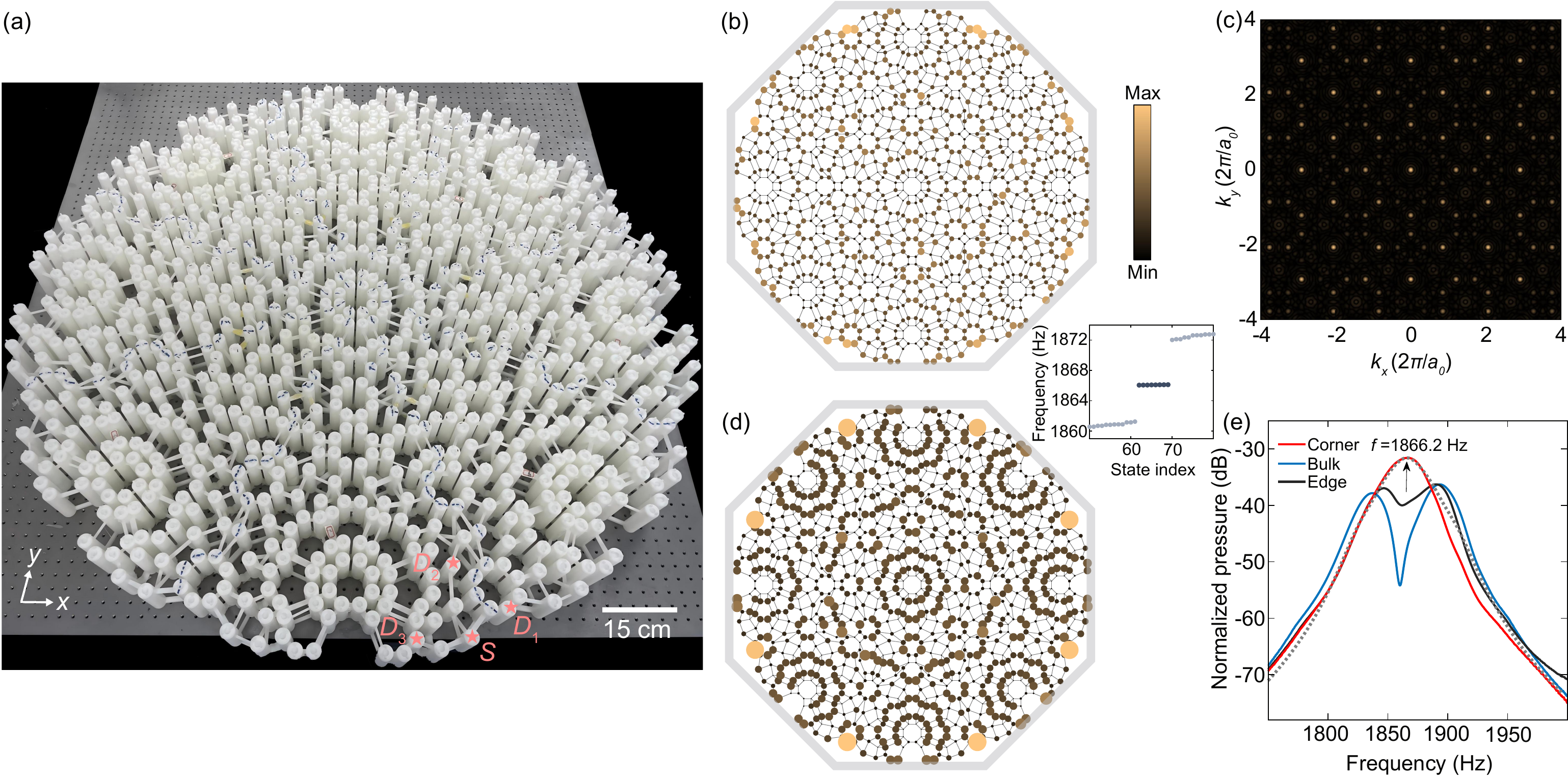}
	\caption{{Observation of the topological phase in an acoustic quasicrystal.} 
		(a) Photograph of the fabricated acoustic quasicrystal sample. 
		(b) Spatial distribution of the measured acoustic pressure field, 
		summed over the 1832--1842 Hz frequency range.
		(c) Fourier spectrum of (b) in the $k_x$--$k_y$ space, 
		which reveals the quasicrystalline feature of the states.
		(d) Measured spatial distribution of the acoustic pressure at the frequency of $1866.2$ Hz,
		showing strong peaks at eight corners, consistent with the theoretical prediction. 
		Inset: simulated eigenfrequencies of the acoustic quasicrystal versus the state index.
		(e) Measured response spectra as a function of frequency detected at 
		corner $D_1$ (red line), bulk $D_2$ (blue line), and edge $D_3$ (black line) 
		given a source $S$ as highlighted in (a).
		The dashed line describes the simulated result for corner modes considering an acoustic loss $\alpha=0.008$, 
		which closely matches the measured result (red line).} 
	\label{Fig2}
\end{figure*}

We start by introducing a tight-binding model on the AB tiling octagonal quasicrystalline lattice. 
As shown in Fig.~\ref{Fig1}(a), the model is built by
placing a cell---containing four, six, or eight lattice sites---at each vertex of the lattice.
These sites are connected by appropriately defined intra-cell and inter-cell bonds 
(see Appendix B for construction details). The corresponding Hamiltonian reads
\begin{align}
	\label{H_TB}
	{H}=&-\sum_{\langle vm,vn\rangle}t_0(|\bm{r}_{vm}-\bm{r}_{vn}|)\xi_{vm,vn}|v,m\rangle \langle v,n| \\ \nonumber
	&-\sum_{\langle\langle vm,wn\rangle\rangle} t_1(|\bm{r}_{vm}-\bm{r}_{wn}|)\xi_{vm,wn} |v,m\rangle \langle w,n|,
\end{align}
where the first term describes the intra-cell hopping from site $n$ to site $m$ within the
same cell $v$ (the position of site $m$ is $\bm{r}_{vm}$), and the second term corresponds to the inter-cell hopping
from site $n$ of cell $w$ to site $m$ of cell $v$.
The hopping exists only between sites connected by bonds represented by 
$\langle \dots \rangle$ (intra-cell) and $\langle\langle \dots \rangle\rangle$ (inter-cell), 
and $t_{j}(r)=t_{j} e^{-\lambda(r-R_{j})}$ ($j=0,1$) is the strength of the corresponding hopping 
which decays exponentially with the distance $r$.  
In addition, we reverse the sign of certain hopping terms [red lines in Fig.~\ref{Fig1}(a)] and increase 
the strength of some hoppings to open the bulk energy gap. These modifications 
are incorporated in the coefficients $\xi_{vm,wn}$ (see Appendix B).
For simplicity, we choose $t_1=1$ and the distance between the central 
positions of adjacent cells ($a_0=1$) as the units of energy and length, respectively. 
Without loss of generality, we set $R_0=0.153$, $R_1=0.630$, and $\lambda=2$
so that the maximum intra-cell and inter-cell hoppings are $t_0$ and $t_1$,
respectively.

This Hamiltonian respects the eight-fold rotational symmetry,
$U_{C_8}HU_{C_8}^{-1}=H$,
and time-reversal symmetry, $THT^{-1}=H$,
where $T=\kappa$ is the complex conjugate operator.  
Since every polygon enclosed by bonds is even-sided, the lattice is 
bipartite, allowing us to partition it into two sublattices
where each bond connects different sublattices. Consequently,
the system possesses chiral (sublattice) symmetry $\Gamma$,
$\Gamma H \Gamma^{-1}=-H$, where $\Gamma$ is a diagonal matrix whose 
entries are $+1$ on one sublattice and $-1$ on the other.

\begin{figure*}[tp]
	\includegraphics[width=1\linewidth]{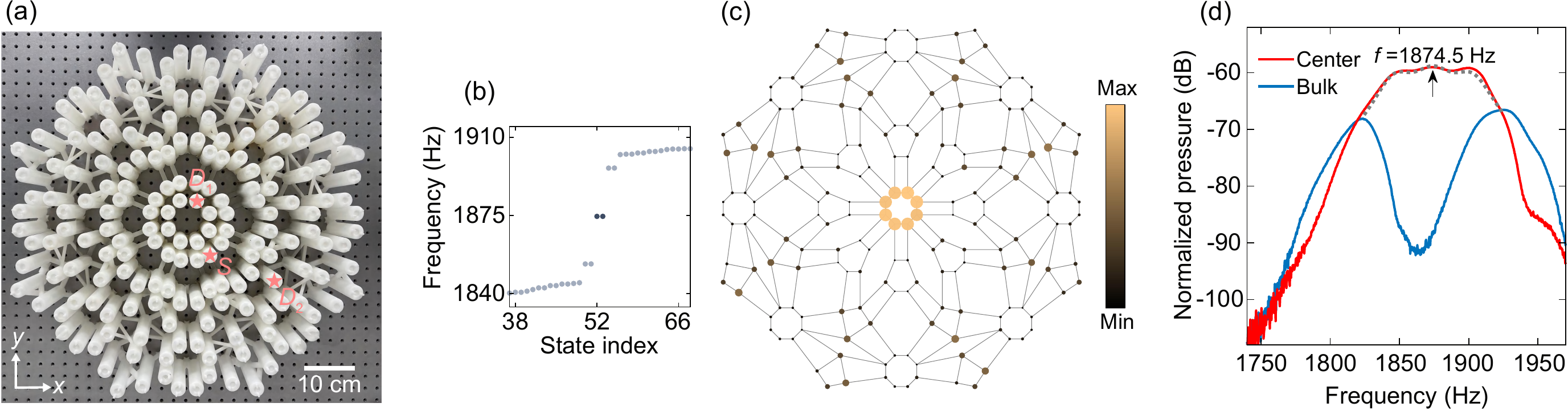}
	\caption{{Observation of the zero-energy modes in a topologically trivial acoustic quasicrystal.} 
		(a) Photograph of the fabricated acoustic quasicrystal sample in the trivial regime 
		(see Appendix J for construction details). 
		(b) Simulated eigenfrequencies of the trivial quasicrystal versus the state index,
		displaying two degenerate states at the frequency of 1874.5 Hz within the energy gap. 
		Note that only the middle pair (dark-blue-filled circles) are zero-energy states (see Appendix G).
		(c) Measured spatial distribution of the acoustic pressure at the frequency of 1874.5 Hz, 
		showing strong peaks at the central cell, which agrees well with the theoretical results. 
		(d) Response spectra as a function of frequency measured at a central position $D_1$ (red line) and 
		a position away from the central cell $D_2$ (blue line)
		given a source $S$ as highlighted in (a).
		The measurement results agree well with the simulated one at the central cell by including 
		an acoustic loss $\alpha=0.008$ (dashed line).} 
	\label{Fig3}
\end{figure*}

To demonstrate the existence of corner modes, 
we numerically compute the eigenenergies and eigenstates of the Hamiltonian and 
display the energy spectrum as a function of $t_0$ in Fig.~\ref{Fig1}(b). 
We see that, when $0<t_0<0.67$, eight corner modes emerge at zero energy inside the bulk gap 
[see the inset of Fig.~\ref{Fig1}(b)]. To visualize their spatial distribution, 
we calculate the local density of states (DOS) defined as 
$\rho(E,\bm{r}_{vm})=\sum_{i}\delta(E-E_i)|\Psi_i(\bm{r}_{vm})|^2$, where 
$\Psi_i(\bm{r}_{vm})$ is the $i$th eigenstate corresponding to the eigenenergy of $E_i$. 
The results illustrate that these modes are mainly localized at the corners with 
polar angles $\theta=\pi (1+2j)/8$ ($j=0,\dots,7$) [see the red circles in Fig.~\ref{Fig1}(a)].
As we further increase $t_0$, the system enters a gapless regime. 
However, for $t_0>4.15$, the bulk gap reopens. Remarkably, two 
zero-energy modes persist in this topologically trivial region.
Unlike the corner modes, these zero-energy modes are
localized at the central cell, as shown in Fig.~\ref{Fig3}(c).
This is in stark contrast to crystalline systems, where central modes are topologically nontrivial 
and arise from the introduction of geometric defects, such as disclinations~\cite{TeoPRL2013}.
We prove that such modes arise in a system  
with eight-fold rotational, 
time-reversal and chiral symmetries, provided that there are an odd number of
sites in an octant (see Appendix C).
For large $t_0$, this behavior can also be understood: Since intra-cell hoppings 
dominate, neglecting the inter-cell hoppings leads to the eigenenergies 
$\{-2t_0, -\sqrt{2} t_0,-\sqrt{2} t_0,0,0,\sqrt{2} t_0, \sqrt{2} t_0, 2t_0\}$ 
so that the central cell hosts a pair of zero-energy modes. 
See also Appendix D for results of a larger
system.

\section{Topological characterization}
We now develop a bulk topological invariant to characterize this 
topological phase. We first introduce twisted boundary conditions 
on the octagonal quasicrystalline structure and construct a 
momentum-space Hamiltonian $H(k)$ with $k\in[0,2\pi]$ (see Appendix E). 
Then, we define a topological invariant based on $H(k)$ at 
high-symmetry points (HSPs) $K=0$ and $\pi$, which commutes 
with ${{U}_{C_8}}$, i.e., $[{U}_{C_8},H(K)]=0$~\cite{ChenPRB2012,ChenPRB2013,TeoPRL2013,BenalcazarPRB2014,GuidoPRB2018,BenalcazarPRB2019}. 
Given that $({U}_{C_8})^8=1$, the rotation eigenvalues are 
$\omega_p=e^{i\pi p/4}$ with $p=0,\dots,7$, and the eigenspace 
corresponding to each $\omega_p$ is generated by the eigenvectors 
$|\omega_{p,q} \rangle$ with $q=1,\dots,N/8$, where $N$ is the 
number of lattice sites. We restrict the Hamiltonian to 
the $\omega_p$ eigenspace, denoted by $H_p(K)$. The restricted 
Hamiltonian belongs to class A in zero dimension, and thus a 
$\mathbb{Z}$ invariant $N_{K}^{(p)}$ can be defined as the 
number of occupied states below zero energy. In conventional 
crystalline materials, trivial atomic insulators are characterized 
by identical values of $N_{K}^{(p)}$ at different HSPs within 
each $\omega_p$ eigenspace, whereas nontrivial topological 
insulators exhibit variations in $N_{K}^{(p)}$ between the 
HSPs~\cite{BenalcazarPRB2014,BenalcazarPRB2019}. Inspired by 
this, we define a topological invariant for each eigenspace 
as the difference in $N_{K}^{(p)}$ between the HSPs, namely, 
\begin{equation}
	\label{chi}
	\chi_p=N_{\pi}^{(p)}-N_{0}^{(p)},
\end{equation}
with $p=0,\dots,7$ (see Appendix E for a discussion of why the 
topological invariant can diagnose the zero-energy corner modes). 
In fact, some constraints reduce the number of independent $\chi_p$. 
First, due to time-reversal symmetry, the eigenspaces of $\omega_p$ and 
$\omega_{8-p}$ are related by complex conjugation, and it follows that 
$\chi_p=\chi_{8-p}$ for $p=1,2,3$. Additionally, the conservation of 
the total number of occupied states ensures $\sum_{p}\chi_p=0$.
In Appendix F, we consider the periodic approximant of our quasicrystalline model~\cite{jagannathan2024properties}. 
In the nontrivial regime with small $t_0$, 
the approximant is gapped and hosts neither gapless edge modes nor corner modes. 
Similarly, based on the four-fold ($C_4$) and two-fold ($C_2$) rotational symmetries of the periodic approximant, 
we define the corresponding $C_4$ topological invariants $\chi_{4,m}$ ($m=0,\dots,3$) 
and $C_2$ topological invariants $\chi_{2,n}$ ($n=0,1$) at the HSPs of its momentum-space Hamiltonian. 
We find that all $\chi_{4,m}$ and $\chi_{2,n}$ vanish, which is consistent with the absence of corner modes.
This indicates the crucial role of eight-fold rotational symmetry in supporting the corner modes in our model.

The topological invariant $\chi_p$, plotted as function of 
$t_0$ in Fig.~\ref{Fig1}(c),
shows that corner modes emerge in the regime with nonzero $\chi_p$, 
indicating their topological origin. 
In this regime, $\chi_p$ take quantized values of $\pm N_b/8$ with $N_b$ being the number of boundary sites. 
Conversely, $\chi_p$ vanish in the bulk gapped regime for $t_0>4.15$, 
indicating that the zero-energy modes localized at the central cell appear in
a topologically trivial phase.

\section{Acoustic experimental realization}
We now experimentally realize the topological quasicrystalline model using acoustic resonators.
As shown in Fig.~\ref{Fig2}(a), we design and fabricate an acoustic quasicrystalline sample. 
The sample is composed of unified acoustic cavities connected by thin tubes mimicking the structure shown in Fig.~\ref{Fig1}(a). 
Each cylindrical cavity with diameter $d=2.20$ cm and height $h=9.25$ cm 
acts as a lattice site with a dipole resonance mode at frequency $f=1870$ Hz. 
These cavities are coupled by various tubes, with the strength and sign of the coupling determined by 
the geometry and connection configuration of tubes~\cite{ni2020demonstration,Qi2020PRL}. The specific arrangement shown in 
Fig.~\ref{Fig2}(a) realizes the tight-binding Hamiltonian in Eq.~\ref{H_TB} (see Appendix G).  

To show that the states in the acoustic sample exhibit a quasicrystalline structure, 
we measure the spatial distribution of acoustic pressure contributed by states near a 
specified frequency. Specifically, we put a headphone at the bottom of a resonator 
as a point source to excite the acoustic signal at a particular frequency and insert a probe with the 
diameter of less than 3 mm into the resonator as a signal detector to record the acoustic 
signal. We excite each resonator at position $\bm{r}_{vm}$ and measure the acoustic pressure at that resonator.
Experimentally, the spatial distribution of the acoustic pressure field, summed over the 1832--1842 Hz 
frequency range and denoted as $P(\bm{r}_{vm})$, is obtained and illustrated in Fig.~\ref{Fig2}(b). 
We further perform a discrete Fourier transform based on its cell structure, defined as 
$F(\bm{k})=\sum_{vm}P(\bm{r}_{vm})e^{-i\bm{k}\cdot\bm{r}_v}$, 
and the resulting Fourier spectrum in $k_x$--$k_y$ space [see Fig.~\ref{Fig2}(c)] exhibits the eight-fold rotational symmetry,
closely resembling that of the octagonal AB tiling quasicrystal~\cite{Kuo1987PRL} 
(see Appendix H). 

To confirm the presence of corner modes in the acoustic quasicrystal, we conduct numerical 
simulations of the sample's acoustic eigenfrequencies using the commercial COMSOL 
Multiphysics solver package. 
The simulated eigenfrequencies, shown in the inset of Fig.~\ref{Fig2}(d), reveal the presence of eight corner 
modes at a frequency of $f=1866.2$ Hz (dark blue circles) that emerge within the gap.  
Experimentally, the measured acoustic pressure field near this frequency 
displays sharp peaks at the corners [see Fig.~\ref{Fig2}(d)], aligning with the theoretical prediction of localized corner modes. 
In addition, the experimentally measured response spectra with respect to the frequency 
at corner $D_1$, edge $D_2$, and bulk $D_3$, following excitation at source $S$,   
are presented in Fig.~\ref{Fig2}(e). We observe that, unlike the dip in the edge and bulk spectra, 
the corner spectrum shows a relatively broad peak near the frequency of the corner modes, consistent with 
the simulated eigenfrequencies. To further validate the measured result, we simulate the response spectrum 
at the corner using a complex acoustic velocity $v=345\times(1+\alpha i)$ m/s, where the imaginary part 
represents propagation loss and $\alpha$ is the loss coefficient. The result for $\alpha=0.008$ 
(dashed line in Fig.~\ref{Fig2}e) shows a good agreement with the experimental data (see Appendix I).

To reveal the zero-energy localized modes in the trivial regime of the quasicrystal system, 
we fabricate a different acoustic quasicrystal, as shown in Fig.~\ref{Fig3}(a). 
The simulated eigenfrequencies of this structure [see Fig.~\ref{Fig3}(b)] confirm the 
presence of two isolated modes within the gap. Experimentally, we measure the acoustic 
pressure distribution at $f=1874.5$ Hz, showing that the distribution is primarily localized at the central cell 
[see Fig.~\ref{Fig3}(c)], which aligns well with the theoretical and simulated results. 
In addition, we measure the response spectra at the central cell $D_1$ and bulk cell $D_2$ 
away from the center with the excitation at source $S$ [as indicated in Fig.~\ref{Fig3}(a)]. 
We see that the central-cell spectrum displays a broad peak at the corresponding eigenfrequency, 
in contrast to the dip in the $D_2$ spectrum. Similar to the topological case in Fig.~\ref{Fig2}(e), 
the simulated central-cell spectrum with a loss coefficient $\alpha=0.008$ [dashed line in Fig.~\ref{Fig3}(d)] 
is consistent with the experimental result (see Appendix I).

\section{Conclusion}
This work provides the experimental observation of topological phases without crystalline
counterparts in an acoustic quasicrystal.
It is based on a novel method for constructing tight-binding models that 
support higher-order topological phases without crystalline counterparts. 
The topological phase is robust against weak disorder as shown in Appendix L.
Building on the design principle, we also propose a decagonal 
quasicrystalline structure, which demonstrates a topologically nontrivial phase featuring 
ten zero-energy corner modes (see Appendix M). 
In addition to acoustic lattices, 
these models can be realized in other metamaterials, such as photonic crystals~\cite{freedman2006wave,wang2020localization,wang2009observation,zilberberg2018photonic,noh2018topological,yang2019realization,li2023exceptional,liu2025photonic}, 
interacting gyroscopes~\cite{mitchell2018amorphous}, and electric circuits~\cite{lee2018topolectrical,imhof2018topolectrical,yang2024circuit}.
By removing the sign-reversal modification, 
the corner states become bound states in the continuum~\cite{Rechtsman2020PRLBound,Benalcazar2020PRB,Xu2020PRB}, which can also be experimentally 
confirmed. 
Including disorder allows for the study of disorder-induced topological phenomena~\cite{Shen2009PRL,Yang2020PRB,Shen2020PRL,Xiong2024PRB}. 
Another promising direction involves extending the structure to three dimensions to realize 
Dirac-like or Weyl-like semimetal phases and three-dimensional topological insulators that lack crystalline 
counterparts~\cite{Donghui2023PRB,mao2024higher}.

\emph{Acknowledgements}---We thank Yan-Bin Yang for helpful discussions.
This work is supported by the Natural Science Foundation of Henan Province (Grant No. 242300421160), 
the National Key R \& D Program of China (Grant No. 2022YFA1404500), the Cross-disciplinary Innovative Research Group Project of Henan Province (Grant No. 232300421004), the National Natural Science Foundation of China (Grants No. 12622413, 12125406, 12374360, 12574423, 12474265, and 11974201).
We also acknowledge the support by center of high performance computing, Tsinghua University.

\section*{Appendix A: Tight-binding model in the previous work}
\label{App_A}
\setcounter{equation}{0} \setcounter{figure}{0} \setcounter{table}{0}
\renewcommand{\theequation}{A\arabic{equation}} \renewcommand{\thefigure}{A\arabic{figure}}
\renewcommand{\theHfigure}{A\arabic{figure}}
\renewcommand{\bibnumfmt}[1]{[#1]} \renewcommand{\citenumfont}[1]{#1}

The topological phases without crystalline counterparts in quasicrystals 
are predicted in Refs.~\cite{Fulga2019PRL,Xu2020PRL}.
In their models, four internal degrees of freedom described by two sets of 
Pauli matrices $\{\sigma_j\}$ and $\{\tau_j\}$ are involved at each 
site on the Ammann-Beenker (AB) tiling octagonal quasicrystalline lattice.
Hopping occurs between two sites $j$ and $l$ when they
are linked by an edge denoted by $\langle j,l \rangle$.  
The tight-binding Hamiltonian in Ref.~\cite{Fulga2019PRL} is given by
\begin{equation} \label{H_VPAPF}
	H_{\mathrm{Ref} }=\sum_j {\Psi}_j^\dagger H_j {\Psi}_j
	+\sum_{\langle j,l\rangle} {\Psi}_j^\dagger H_{jl} {\Psi}_l,
\end{equation}
where
${\Psi}_j^\dagger=
\left(\begin{array}{cccc}
	|j,1\rangle & |j,2\rangle & |j,3\rangle & |j,4\rangle \end{array}\right)$ 
with $|j,\nu\rangle$ ($\nu=1,2,3,4$) representing a degree of freedom
at site $j$ on the quasicrystal. 
The mass term is $H_j=\mu\sigma_z\tau_z$, and the hopping matrix is
\begin{align} 
	H_{jl}=&(t/2)\sigma_z\tau_z-(i\Delta/2) \left(\cos \theta_{jl} \sigma_z\tau_x+\sin\theta_{jl} \sigma_z\tau_y \right) \\ \nonumber
	&+(V/2)\sigma_y\tau_0\cos 4\theta_{jl},
\end{align}
where $\mu$, $\Delta$, and $V$ are real parameters
and $\theta_{jl}$ is the polar angle between sites $j$ and $l$.
Note that this Hamiltonian in Eq.~\ref{H_VPAPF} corresponds to the 
Bogoliubov-de-Gennes (BdG) Hamiltonian in Ref.~\cite{Fulga2019PRL}.
Its topology can be characterized by a Pfaffian invariant~\cite{Fulga2019PRL} or a 
generalized quadrupole moment~\cite{tao2023average,mao2024higher}.

\begin{figure}[htp]
	\includegraphics[width=1\linewidth]{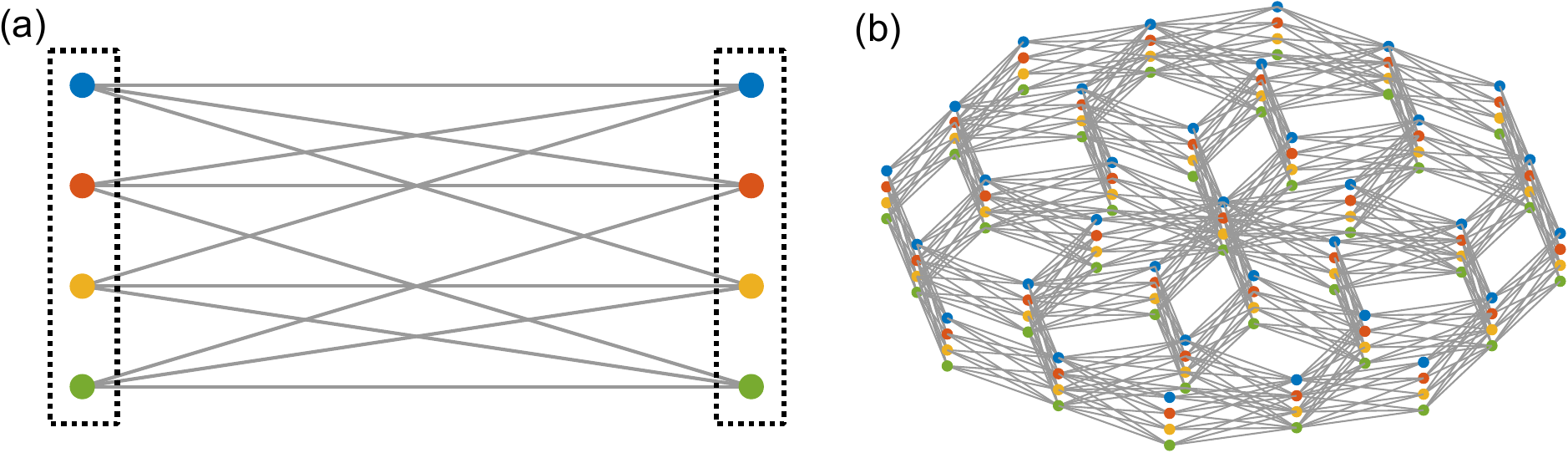}
	\caption{{Illustration of the tight-binding model from Ref.~\cite{Fulga2019PRL}.} 
		(a) Hoppings between two cells, each containing four 
		degrees of freedom.  
		(b) Illustration of the tight-binding model from Ref.~\cite{Fulga2019PRL}
		on an AB tiling octagonal quasicrystalline lattice. Here the grey lines only
		indicate the presence of hoppings, but do not distinguish between  
		real and imaginary values.
	} 
	\label{FigSM_ori_TB}
\end{figure}

To realize the model experimentally in metamaterials (similarly for the model in Ref.~\cite{Xu2020PRL}), 
we need to encode each internal 
degree of freedom into a node, such as a resonator, ensuring that each cell contains four nodes. 
The Hamiltonian in Eq.~\ref{H_VPAPF} requires each node in a cell to 
connect with three nodes in a neighboring cell, as shown in Fig.~\ref{FigSM_ori_TB}(a). 
Arranging all cells into a quasicrystalline structure, as shown in Fig.~\ref{FigSM_ori_TB}(b), 
results in a very complex network, posing significant challenges for experimental realization.
This difficulty is exacerbated by the presence of both real and imaginary hopping amplitudes.

\section*{Appendix B: Construction of a tight-binding model on an octagonal quasicrystalline lattice}
\setcounter{equation}{0} \setcounter{figure}{0} \setcounter{table}{0}
\renewcommand{\theequation}{B\arabic{equation}} \renewcommand{\thefigure}{B\arabic{figure}}
\renewcommand{\theHfigure}{B\arabic{figure}}
\renewcommand{\bibnumfmt}[1]{[#1]} \renewcommand{\citenumfont}[1]{#1}

Here, we will elaborate on how to construct the tight-binding model on an AB tiling octagonal quasicrystalline lattice.

\begin{figure*}[htp]
	\centering
	\includegraphics[width=0.7\linewidth]{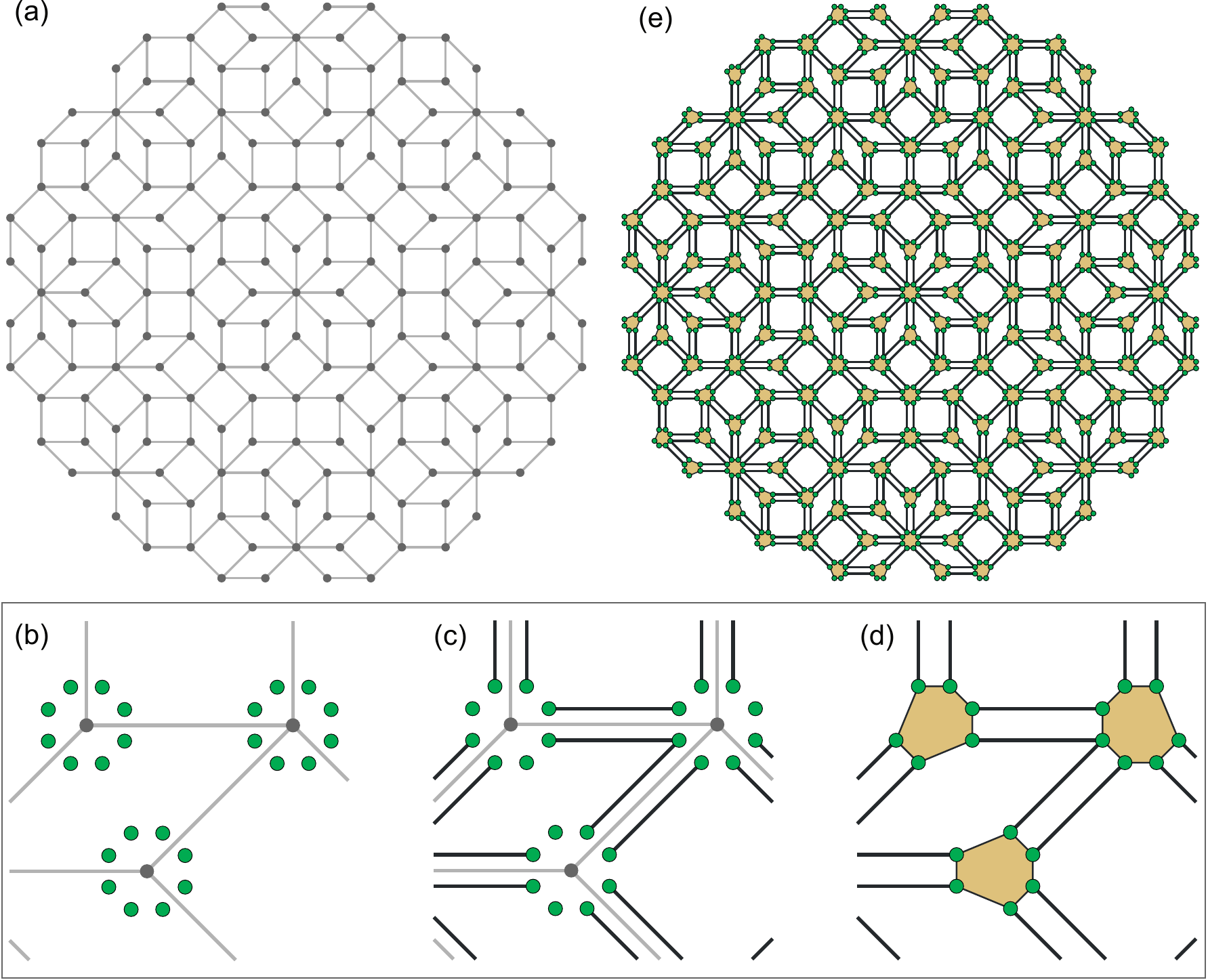}
	\caption{{Construction of the tight-binding model on an octagonal quasicrystalline lattice.} 
		(a) An AB tiling octagonal quasicrystalline lattice. 
		(b)--(d) We first place a cell consisting of eight sites at each vertex [see (b)]
		followed by introducing two edges connecting these sites [see (c)]. 
		In (d), all sites lacking inter-cell bonds are 
		removed and the remaining sites within each cell are connected. 
		(e) Illustration of the obtained tight-binding model. Here, we set $R_c=0.2$.} 
	\label{FigSM_C8_incipient_lattice}
\end{figure*}

We first generate the quasicrystalline lattice [see Fig.~\ref{FigSM_C8_incipient_lattice}(a)] 
via the cut-and-project method~\cite{SMJanot1994,SMLedermann2010,SMcryst7100304}. In this lattice, 
all edges are of uniform length, which we set to unity. 
Second, inspired by primitive generators of topological crystalline insulators~\cite{ChenPRB2012,ChenPRB2013,TeoPRL2013,BenalcazarPRB2014,GuidoPRB2018,BenalcazarPRB2019}, 
we decorate each vertex with a cell containing eight sites. 
These sites are uniformly distributed on a circle of radius $R_c=0.2$ 
centered at a vertex, as illustrated in Fig.~\ref{FigSM_C8_incipient_lattice}(b).
Their positions are given by $\bm{r}_{vm}=\bm{r}_v+\bm{R}_m$, 
where $\bm{r}_v$ is the position of vertex $v$, 
and $\bm{R}_m=R_c(\cos\theta_m,\sin\theta_m)$ with $\theta_m=(2m-1)\pi/8$
for $m=1,\dots,8$. 
Inter-cell hoppings are restricted to pairs of cells connected 
by the underlying quasicrystalline edges. 
For each edge, we establish a pair of parallel, minimal-length 
inter-cell bonds, as depicted in Fig.~\ref{FigSM_C8_incipient_lattice}(c).
Subsequently, we remove all isolated sites that lack inter-cell connections 
and introduce intra-cell bonds between the remaining sites within 
each cell [Fig.~\ref{FigSM_C8_incipient_lattice}(d)]. 
The cell structures at the boundary vertices are determined by 
considering a larger quasicrystal to ensure consistency. 
The final constructed structure is shown in Fig.~\ref{FigSM_C8_incipient_lattice}(e).

\begin{figure*}[htp]
	\includegraphics[width=0.7\linewidth]{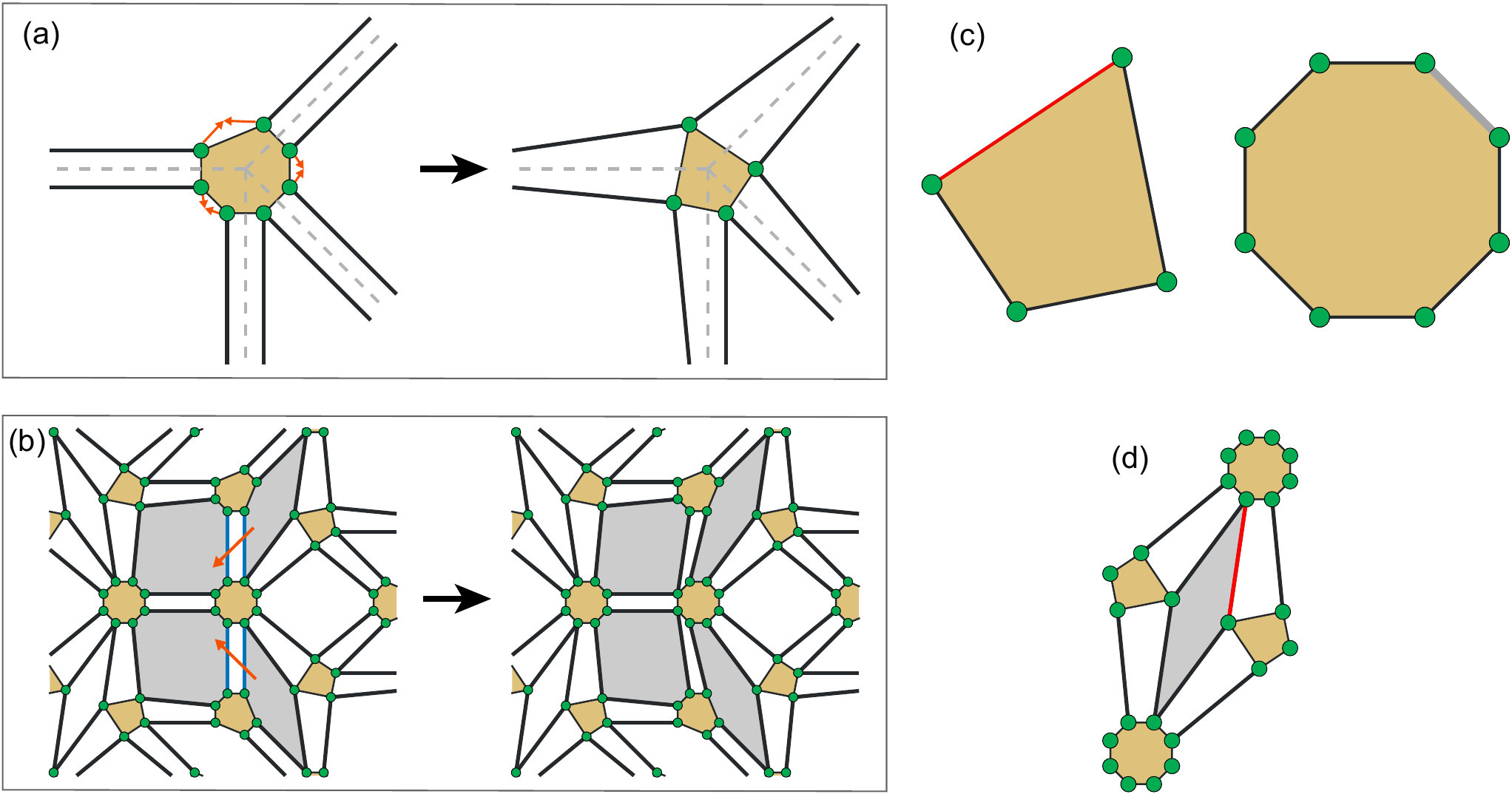}
	\caption{{Modification of certain hoppings opens the bulk energy gap.}
		(a) A cell of seven sites is changed to a cell of four sites. 
		(b) All grey polygons are transformed into even-sided shapes by 
		shifting the blue inter-cell bonds.
		(c) The sign of an intra-cell hopping within each four-site cell is reversed (red line), and the magnitude of an intra-cell hopping 
		in all eight-site cells---except the
		eight-site cell at the center---is increased (grey line).
		(d) The sign of an inter-cell hopping within a slender rhombus is also reversed.
	}
	\label{FigSM_C8_modi}
\end{figure*}

To enforce chiral (sublattice) symmetry, we modify the lattice to ensure that
all polygons enclosed by bonds are even-sided. 
This is achieved through two primary adjustments.
First, for cells containing seven sites, 
if two edges (dashed lines) intersect 
at an angle exceeding $\pi/4$,
the two sites within this angular region are consolidated into a single site, 
as illustrated in Fig.~\ref{FigSM_C8_modi}(a). 
Second, to address odd-sided polygons formed by intra- and inter-cell bonds 
[grey polygons in Fig.~\ref{FigSM_C8_modi}(b)], the connecting inter-cell bonds 
(blue) are shifted. This bond relocation transforms the offending polygons 
into even-sided shapes.

At this stage, the model can support zero-energy corner modes; however, 
its bulk energy spectrum is gapless. Although such modes are of interest as bound states in the continuum~\cite{Rechtsman2020PRLBound,Benalcazar2020PRB,Xu2020PRB}, 
our objective is to realize a system with a gapped bulk spectrum. 
To achieve this, we introduce modifications to specific intra-cell and inter-cell hopping strengths.
For simplicity, we consider two limiting regimes:  
$t_0/t_1\gg 1$ and $t_0/t_1\ll 1$. In the former regime, intra-cell hoppings dominate, 
resulting in zero-energy states within four-site and eight-site cells. To open the gap, 
we reverse the sign of one hopping in the four-site cells and enhance the strength of one hopping 
in the eight-site cells, as indicated by the red and grey lines in Fig.~\ref{FigSM_C8_modi}(c). 
The discussion of the case without modifying the strength is shown in the following Appendix C.
Conversely, in the $t_0/t_1\ll 1$ regime, dominant inter-cell hoppings induce zero-energy states in the slender 
rhombus structure [grey polygon in Fig.~\ref{FigSM_C8_modi}(d)]. 
Here, a gap is opened by reversing the sign of one hopping (red line), 
analogous to the modification for the four-site cells. 
Critically, all modifications are applied in a manner that preserves the system's 
eight-fold rotational symmetry. The resulting octagonal quasicrystalline structure 
is depicted in Fig.~\ref{Fig1}(a).
Note that our construction rules do not apply to random graphs, as these lack any regular local connection rules.

\section*{Appendix C: Zero-energy modes in the trivial case}
\setcounter{equation}{0} \setcounter{figure}{0} \setcounter{table}{0}
\renewcommand{\theequation}{C\arabic{equation}} \renewcommand{\thefigure}{C\arabic{figure}}
\renewcommand{\theHfigure}{C\arabic{figure}}
\renewcommand{\bibnumfmt}[1]{[#1]} \renewcommand{\citenumfont}[1]{#1}

\subsection*{1. Symmetry protection}
We now prove that the two zero-energy modes in the topologically trivial phase
are protected by eight-fold rotational symmetry $U_{C_8}$, time-reversal symmetry $T$, and chiral symmetry $\Gamma$.

We first restrict the Hamiltonian under open boundary conditions to 
the $\omega_p$ eigenspace of $U_{C_8}$ with $p=0,\dots,7$. 
The restricted Hamiltonian is given by $H_p=V_p^\dagger H V_p$, where $V_p$ is a 
$N\times N_s$ matrix defined as $V_p=(|\omega_{p,1}\rangle,\dots,|\omega_{p,N_s} \rangle)$ with 
$|\omega_{p,q} \rangle$ being an eigenvector 
of $U_{C_8}$ with eigenvalue of $\omega_p=e^{i\pi p/4}$
($q=1,\dots,N_s$, $N_s=N/8$, and $N$ is the total number of lattice sites).
The eigenenergies of $H_p$, labeled as $E_{p,h}$ with $h=1,\dots,N_s$, are ordered such that $E_{p,1}\le\dots \le E_{p,N_s}$, 
and the set of these eigenenergies is defined as $S_p\equiv\{E_{p,h}\mid h=1,\dots,N_s\}$.

\begin{figure}[htp]
	\includegraphics[width=0.7\linewidth]{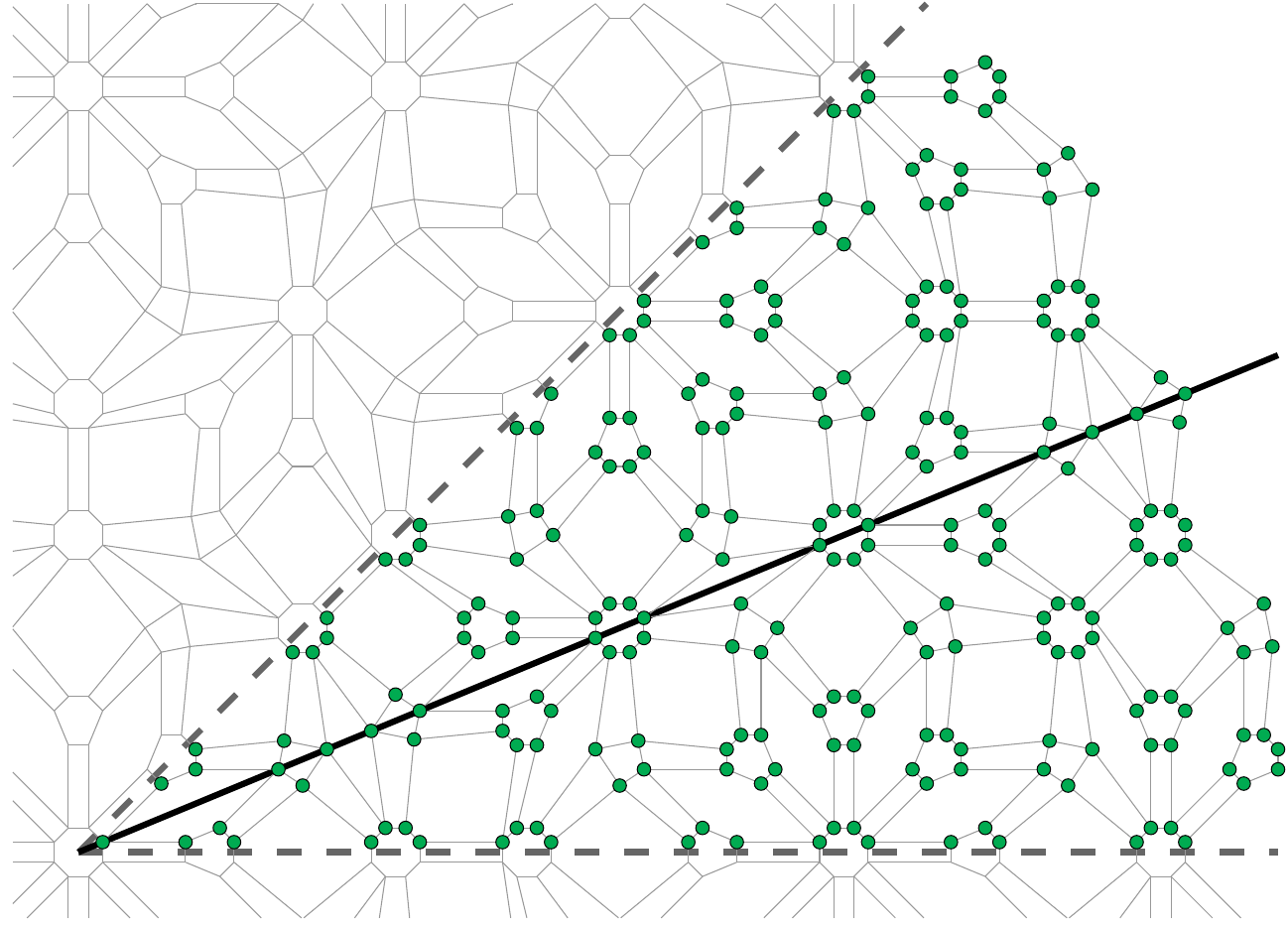}
	\caption{{Illustration of an 1/8 sector of the octagonal quasicrystalline structure.}
		Green circles represent the sites within the 1/8 sector and black line denotes the symmetry axis of the sector. 
		Except for the single site from the central cell, 
		each cell traversed by the symmetry axis provides a pair of sites lying on the symmetry axis. 
		Thus, the number of sites within the sector must be odd. The number of sites in the sector is 239.} 
	\label{FigSM_C8_sector}
\end{figure}

Due to time-reversal symmetry, $S_p$ and $S_{8-p}$ ($p=1,2,3$) are 
related by complex conjugation, and it follows that $S_p=S_{8-p}$.
In addition, based on the relation $\{U_{C_8},\Gamma\}=0$, we apply $\Gamma$ 
on the eigenequation $U_{C_8}|\omega_{p,q}\rangle=\omega_p|\omega_{p,q} \rangle$, leading to
\begin{align}
	\label{SL}
	\Gamma U_{C_8}|\omega_{p,q}\rangle=-U_{C_8}\Gamma|\omega_{p,q}\rangle=\omega_p\Gamma|\omega_{p,q}\rangle.
\end{align}
Since $\omega_{\text{mod}(p+4,8)}=-\omega_p$, $\Gamma$ maps $|\omega_{p,q}\rangle$ to the $\omega_{\text{mod}(p+4,8)}$ eigenspace. 
Thus, chiral symmetry establishes a relation between the eigenspaces of $\omega_{p}$ and $\omega_{\text{mod}(p+4,8)}$. 
Given that $\Gamma H \Gamma^{-1}=-H$, it follows that $S_{\text{mod}(p+4,8)}=-S_p$.

When $p=2$, we have $S_2=S_6=-S_6$. This means that $S_2$ and $S_6$ are the same symmetric sets 
whose elements form positive--negative pairs. If $N_s=2l+1$ ($l$ is a positive integer), then
\begin{align}
	\label{Ns_odd}
	S_2=S_6=\{-E_{l},\dots,-E_{1},0,E_{1},\dots,E_{l}\}
\end{align}
with $0\le E_{1}\le\dots\le E_{l}$. In other words, an odd number of sites in the 1/8 sector guarantees the presence of two zero-energy modes.
We can check that this cannot occur in crystalline systems with two-fold, four-fold, three-fold or six-fold rotational symmetries.

Figure~\ref{FigSM_C8_sector} displays a 1/8 sector of the 
octagonal quasicrystalline structure. 
In fact, the AB tiling octangle quasicrystalline lattice via the cut-and-project method is inherently axisymmetric, 
and the introduction of cell structures preserves this symmetry. Specifically, 
the lattice shown in Fig.~\ref{FigSM_C8_sector} is symmetric about the black line. 
Consequently, the number of sites away from the symmetry axis within the 1/8 sector must be even. On the symmetry axis itself (the black line), 
the central cell contributes a single site, whereas all other traversed cells contribute an even number of sites, 
as each cell contains an even number of sites, as illustrated in Fig.~\ref{FigSM_C8_sector}. 
Thus, each 1/8 sector contains an odd number of sites regardless of system size, necessitating the presence of two zero-energy modes.
In the topologically nontrivial regime, these modes are localized at the corners; 
in the trivial regime, they are localized at the center; in the gapless regime, 
they appear at other positions.

\begin{figure*}[htp]
	\includegraphics[width=0.8\linewidth]{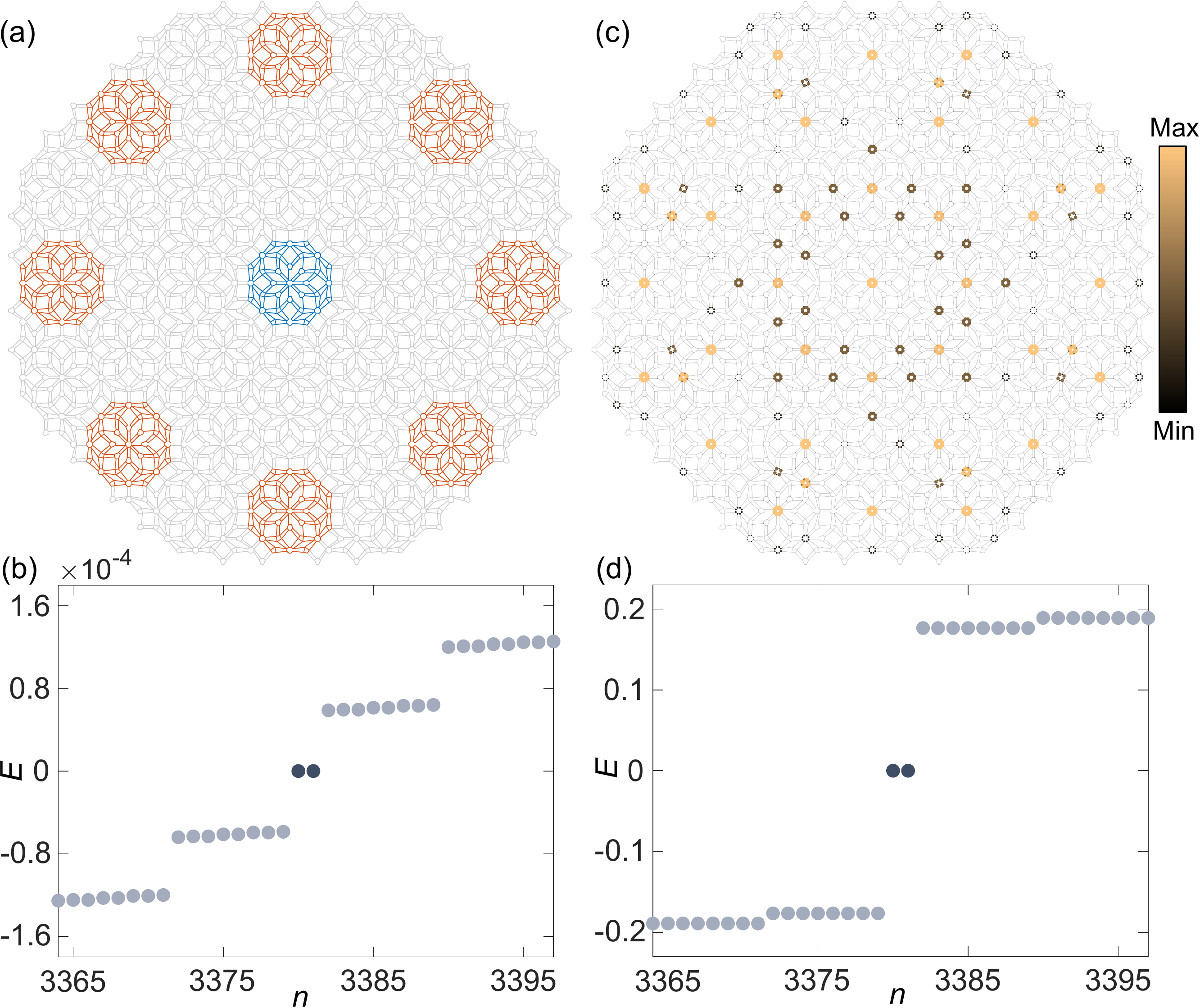}
	\caption{{Repetitivity of the quasicrystalline structure in a larger scale.} 
		(a) Geometric structure of the octagonal quasicrystalline structure. 
		Blue lines describe the central patch corresponding to the trivial case in Fig.~\ref{Fig3}, 
		and red lines depict the repeating patches with the same geometric structure as the central one. 
		The number of sites in each patch is 184 and the total number of sites for the larger structure is 6760. 
		(b) Energy spectrum versus the state index $n$ and 
		(c) spatial distribution of the local DOS at zero energy 
		for the tight-binding Hamiltonian constructed by (a) without the modification of eight-site cells. 
		(d) Energy spectrum versus the state index $n$ for the modified Hamiltonian with the same size in (a). 
		Here, we set $t_0=10$ and all other system parameters are the same as those in Fig.~\ref{Fig1}.
	}
	\label{Fig_reply_repeat}
\end{figure*}

\subsection*{2. The unmodified case}
As shown in Appendix C, to open an energy gap in the bulk, we have slightly modified an intra-cell hopping 
in all eight-site cells except the eight-site cell at the center. This adjustment preserves rotational symmetry 
but results in distinct local Hamiltonians for the central patch (blue lines) and the repeating patches (red lines) 
shown in Fig.~\ref{Fig_reply_repeat}(a). 
Consequently, in the trivial phase, only the central patch contributes two exact zero-energy modes, 
while the repeating patches do not.

First, consider the unmodified case where the intra-cell hopping within the eight-site cells is unchanged. 
The energy spectrum shown in Fig.~\ref{Fig_reply_repeat}(b) reveals that, 
a pair of exact zero-energy modes (dark-blue-filled circles) still exist, as protected by symmetries. 
Yet, numerous near-zero-energy modes appear. 
Some of these near-zero modes are localized within the repeating patches, 
while others are confined to other eight-site cells, 
as illustrated by the local DOS at zero energy in Fig.~\ref{Fig_reply_repeat}(c). 
This behavior arises because, when $t_0/t_1\gg 1$, intra-cell hoppings dominate. 
Neglecting the inter-cell hoppings leads to the eigenenergies 
$\{-2t_0,-\sqrt{2} t_0,-\sqrt{2} t_0,0,0,\sqrt{2} t_0,\sqrt{2} t_0,2t_0\}$ 
so that zero-energy modes appear. 
However, due to the coupling between these zero-energy modes, 
a very small minigap is observed.

Since we are interested in a system with a gapped bulk spectrum, 
the magnitude of an intra-cell hopping in all eight-site cells---except the eight-site cell 
at the center---is increased (see Fig.~\ref{FigSM_C8_modi} in Appendix B). 
Note that to preserve the eight-fold rotational symmetry, 
only the central eight-site cell located at the origin remained unmodified, 
corresponding to the unique rotation center. In the trivial regime, 
this results in zero-energy modes localized solely within the central eight-site cell, 
while the spectra of all other patches are gapped. 
See Fig.~\ref{Fig_reply_repeat}(d) for the eigenenergies for this large sample, 
confirming the existence of two zero-energy modes at the center.

\begin{figure*}[htp]
	\includegraphics[width=1\linewidth]{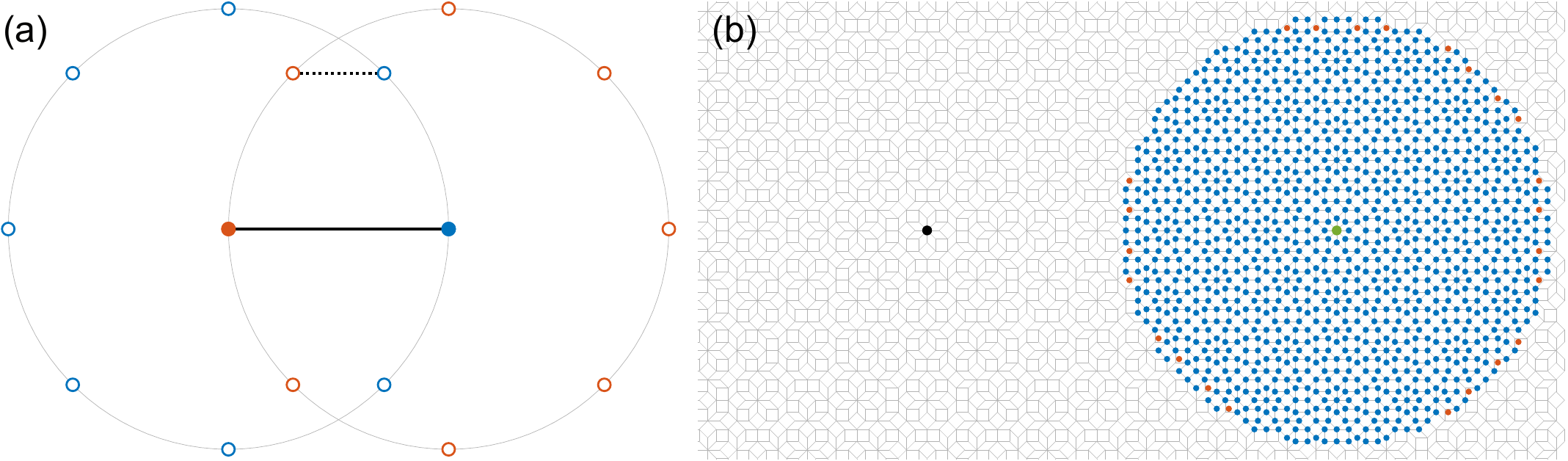}
	\caption{{Sole strict rotation center in the octagonal quasicrystal.} 
		(a) Contradiction for a pair of rotation centers with minimal separation in eight-fold rotational symmetry. 
		Red- and blue-filled circles represent a pair of rotation centers with assumed minimal separation 
		and other colored hollow points are generated by applying $C_8$ rotations about each center. 
		Clearly, the black dotted line is  shorter than the black solid one. 
		(b) Illustration of the approximate eight-fold rotational symmetry in the AB tiling octangle quasicrystal. 
		Black- and green-filled circles denote the strict and approximate rotation centers, respectively. 
		Red- and blue-filled circles denote the configuration upon rotating the lattice by $\pi/4$ 
		about the approximate center. Blue circles coincide with original lattice sites, while red circles do not.
	} 
	\label{Fig_reply_repeat_center_unique}
\end{figure*}

In addition, we would to clarify that there exists only one strict rotation center,
although quasicrystals exhibit local repetitivity~\cite{baake2012notions}. 
Specifically, a system on a Euclidean space with $n$-fold rotational symmetry 
can possess only one rotation center when $n=5$ or $n\ge7$. 
Considering $n=8$, suppose the system had more than one distinct rotation center. 
We choose the pair with minimal separation (solid black line), marked by red-filled and blue-filled circles 
in Fig.~\ref{Fig_reply_repeat_center_unique}(a). 
Applying $C_8$ rotations about each center would generate new centers (hollow circles). 
Among these, a pair with a smaller separation (black dotted line) would inevitably appear, 
implying an infinitely reducible minimal distance---a contradiction. 
This uniqueness is also apparent in the AB tiling octagonal quasicrystalline lattice 
[see Fig.~\ref{Fig_reply_repeat_center_unique}(b)], 
where the sole strict rotation center is marked by a black-filled circle. 
Due to the quasicrystal's repetitivity, an approximate rotation center (green-filled circle) can be identified. 
However, rotating the lattice by $\pi/4$ about this approximate center produces a configuration 
(red- and blue-filled circles) where the red circles do not coincide with original lattice sites, 
confirming the breaking of exact rotational symmetry away from the true center.

\subsection*{3. The choice of modification pattern}
\begin{figure*}[htp]
	\includegraphics[width=0.8\linewidth]{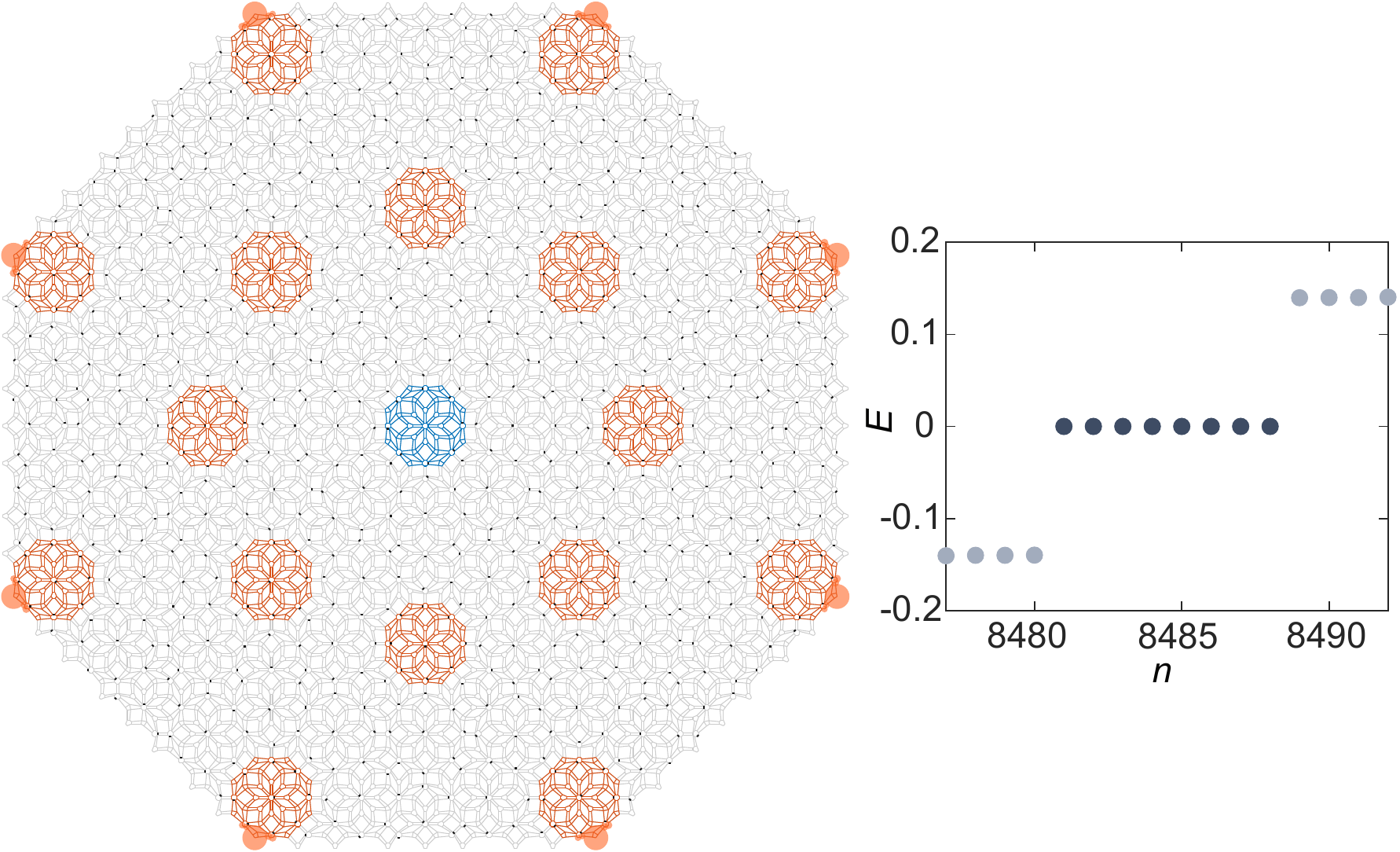}
	\caption{Quasicrystalline structure with randomly selected modification patterns for the eight-site cells.
			Left panel: Geometric structure of the octagonal quasicrystalline structure with central (blue lines) and repeating (red lines) patches. 
			Black lines denote the randomly selected hopping modifications for the eight-site cells, which maintain the eight-fold rotational symmetry. 
			Orange-filled circles near the corners	represent the spatial distribution of the zero-energy local DOS, 
			with the relative size of the circles indicating its magnitude.
			The number of sites in each patch is 184 and the total number of sites for the larger structure is 16968. 
			Right panel: the energy spectrum at $t_0 = 0.3$ with respect to the state index $n$, 
			highlighting the presence of eight zero-energy states. 
			Here, all system parameters are the same as those in Fig.~\ref{Fig1}.}
	\label{Fig_reply_random_modi}
\end{figure*}

As discussed above, our model introduces additional hopping modifications to the underlying AB tiling 
and differs from the Hamiltonian with local symmetry discussed in Ref.~\cite{Fulga2019PRL}.
Nevertheless, the specific choice of modified hoppings is not unique, 
provided that the modification preserves the eight-fold rotational symmetry.  
To examine whether the corner modes depend on a fixed modification pattern, 
we consider a larger system with randomly selected hopping modifications for the eight-site cells (black lines) 
while maintaining the eight-fold rotational symmetry, as shown in Fig.~\ref{Fig_reply_random_modi}. 
With this scheme, except for the central patch (blue lines), 
the local hopping modifications of repeating patches (red lines) no longer follow a fixed octant-dependent pattern, 
making it difficult to directly identify their corresponding octants. 
Importantly, as shown by the spectrum and the corresponding zero-energy local DOS in Fig.~\ref{Fig_reply_random_modi}, 
the zero-energy corner modes persist, indicating that their existence does not rely on a specific modification pattern.
	
\section*{Appendix D: Behavior for a larger system}
\setcounter{equation}{0} \setcounter{figure}{0} \setcounter{table}{0}
\renewcommand{\theequation}{D\arabic{equation}} \renewcommand{\thefigure}{D\arabic{figure}}
\renewcommand{\theHfigure}{D\arabic{figure}}
\renewcommand{\bibnumfmt}[1]{[#1]} \renewcommand{\citenumfont}[1]{#1}

\begin{figure*}[htp]
	\includegraphics[width=1\linewidth]{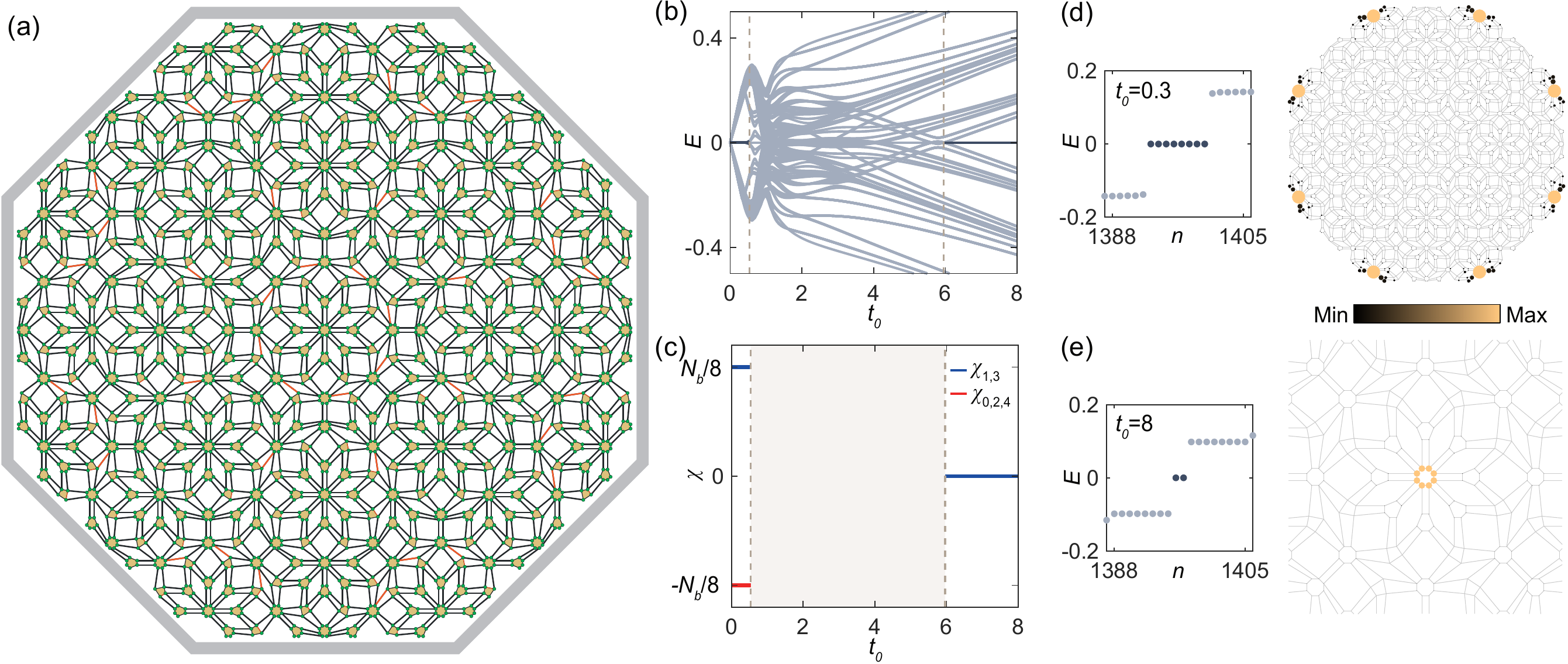}
	\caption{Quasicrystalline structure in a larger scale. 
		(a) Illustration of the tight-binding model on the octagonal quasicrystalline lattice. 
		The total number of sites is $N=2792$. 
		(b) Energy spectrum of the tight-binding Hamiltonian shown in (a) with respect to $t_0$. 
		Similar to the case described in the main text, 
		the system exhibits a topological regime with eight corner modes for $0<t_0<0.54$ 
		and a trivial regime with two zero-energy modes for $t_0>5.96$. 
		(c) Topological invariants $\chi_p$ as a function of $t_0$ with $p=0,\dots,4$. 
		In the topological regime, $\chi_p=\pm N_b/8$ with $N_b=80$, whereas in the trivial regime, $\chi_p=0$. 
		(d)--(e) Energy spectrum versus the state index $n$ and spatial distribution of the local DOS at zero energy 
		for (d) $t_0=0.3$ and (e) $t_0=8$. The size of the circles represents the magnitude of the local DOS, 
		complementing the color representation. Here, all system parameters are the same as those in Fig.~\ref{Fig1}.
	} 
	\label{Fig_reply_larger_size}
\end{figure*}

We now compute the quasicrystalline structure with a larger size $N=2792$, 
and the illustration of the corresponding tight-binding Hamiltonian is shown in Fig.~\ref{Fig_reply_larger_size}(a). 
The resulting phase diagram about the energy spectrum and topological invariants $\chi_p$ as a function of $t_0$, 
shown in Figs.~\ref{Fig_reply_larger_size}(b) and (c), is similar to that in Fig.~\ref{Fig1}, 
which exhibits a topological regime with eight corner modes [see Fig.~\ref{Fig_reply_larger_size}(d)] 
for $0<t_0<0.54$ and a trivial regime with two zero-energy modes [see Fig.~\ref{Fig_reply_larger_size}(e)] for $t_0>5.96$. 

\section*{Appendix E: Construction of the momentum-space Hamiltonian}
\setcounter{equation}{0} \setcounter{figure}{0} \setcounter{table}{0}
\renewcommand{\theequation}{E\arabic{equation}} \renewcommand{\thefigure}{E\arabic{figure}}
\renewcommand{\theHfigure}{E\arabic{figure}}
\renewcommand{\bibnumfmt}[1]{[#1]} \renewcommand{\citenumfont}[1]{#1}

\begin{figure}[htp]
	\includegraphics[width=0.8\linewidth]{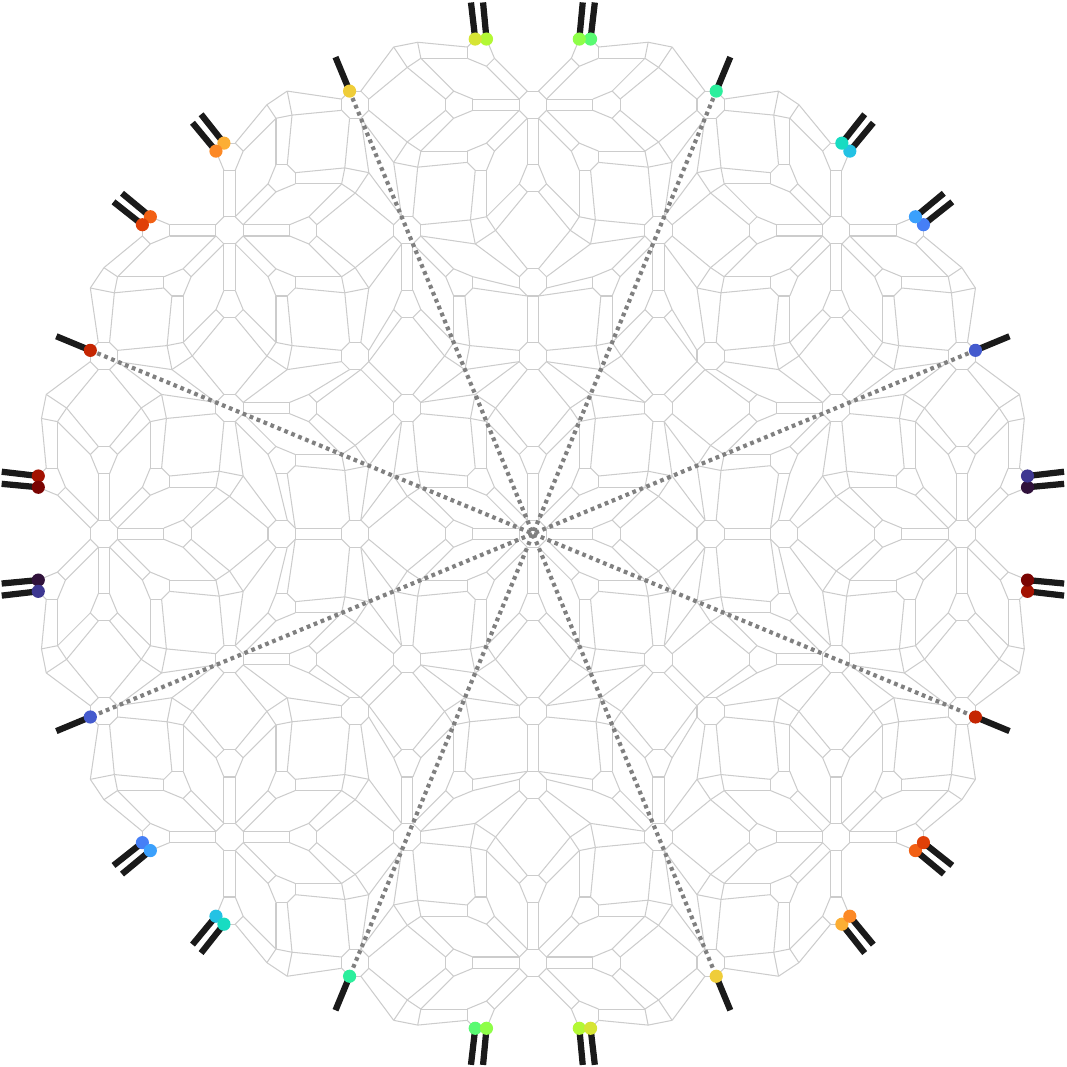}
	\caption{{Boundary sites and boundary hoppings under twisted boundary conditions.} 
		Colored circles denote boundary sites and boundary hopping (black lines) exist 
		between pairs of sites with the same color.} 
	\label{FigSM_C8_PBC}
\end{figure}

To construct a momentum-space Hamiltonian using twisted boundary conditions,
we first identify boundary sites as those lacking inter-cell bonds, 
marked by colored circles in Fig.~\ref{FigSM_C8_PBC}. 
We then impose boundary hoppings (black lines in Fig.~\ref{FigSM_C8_PBC}) 
exclusively between boundary sites of the same color, with the two sites 
in each pair related by inversion symmetry.

For the resulting twisted Hamiltonian, these specific 
boundary hoppings acquire an extra phase factor of $e^{-ik}$,
leading to the boundary term
\begin{align}
	\label{TBC}
	{T}(t_b,k)=-\sum_{\langle\bm{r}_b,I \bm{r}_b\rangle}(t_b e^{-ik}|I\bm{r}_{b}\rangle \langle \bm{r}_{b} |+\text{h.c.}),
\end{align}
where $t_b$ represents the strength of boundary hopping, 
$|\bm{r}_{b}\rangle$ denotes a state at the boundary site located at $\bm{r}_b$, 
$I$ is the spatial inversion operator, 
$\langle \dots \rangle$ denotes a pair of connected boundary sites,
and the hopping strength is set to one. 
Consequently, the momentum-space Hamiltonian is given by ${H}(k)={H}+{T}(t_b,k)$ 
with $k\in[0,2\pi]$ and $t_b=1$.

As mentioned in the main text, the eight-fold rotational operator $U_{C_8}$ 
commutes with the Hamiltonian $H(K)$ at the high-symmetry points $K=0$ and $\pi$, 
and we thus define a topological invariant $\chi_p$ for each eigenspace of $U_{C_8}$ with $p=0,\dots,7$. 
In the trivial phase, $\chi_p=0$; otherwise, nonzero values of $\chi_p$ exist. 
In fact, we can set $k=0$ and modify $t_b$ from 1 to -1. When $t_b$ becomes negative, 
this is equivalent to changing $k$ to $\pi$. In the nontrivial phase, 
during this transition, there exists a $p$ such that the number of occupied states below zero 
in this subspace experience a sudden change as $t_b$ changes sign, 
indicating that the energy gap closes at $t_b=0$. 
The zero-energy states must be localized near boundaries. 
In addition, due to the presence of chiral symmetry and time-reversal symmetry $T$ with $T^2=1$, 
the AZ classification for the system belongs to two-dimensional class BDI 
which is topologically characterized by a trivial group~\cite{chiu2016classification}. 
This indicates that the edge states when $t_b=0$ are not first-order. 
The only possibility is that they are second-order corner states. 
The eight-fold rotational symmetry further enforces the existence of eight-fold corner states, 
as for any corner state $|\psi_c\rangle$, $(U_{C_8})^l|\psi_c\rangle$ 
with $l=0,1,\dots,7$ are all corner states.

\begin{figure}[htp]
	\includegraphics[width=0.9\linewidth]{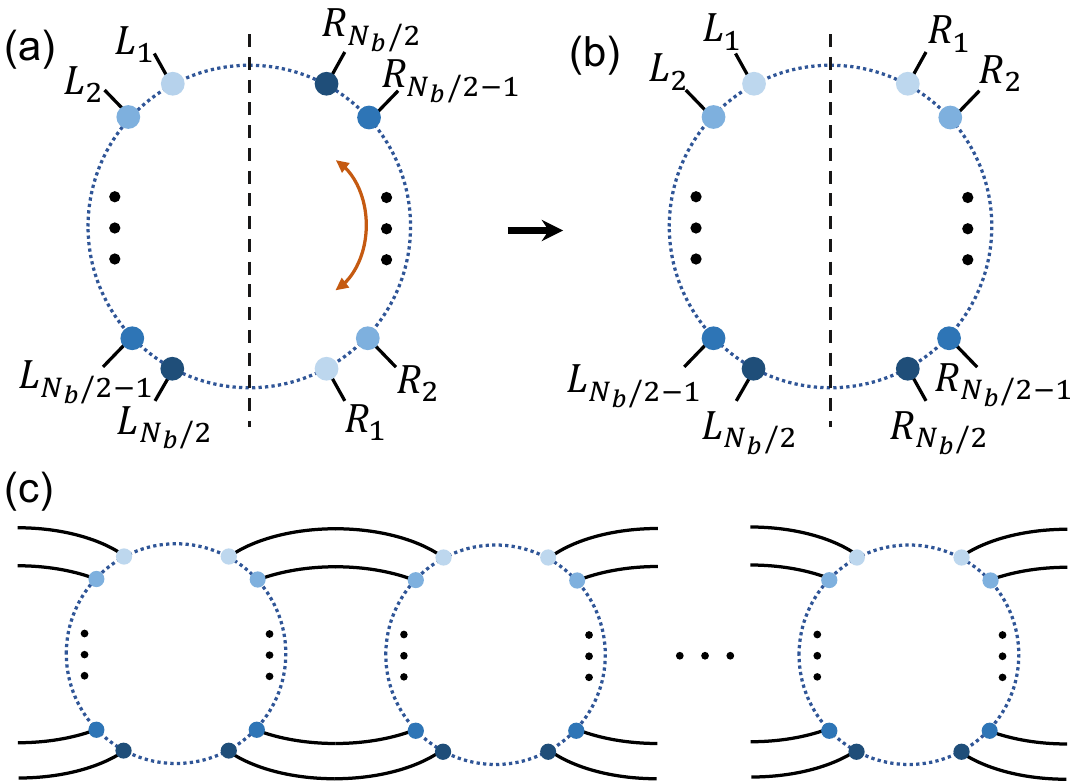}
	\caption{The proposed momentum-space Hamiltonian under twisted boundary 
		conditions corresponds to a one-dimensional periodic chain in which the entire quasicrystalline 
		structure is regarded as a large-scale unit cell.  
		(a)--(b) Illustration of the large-scale unit cell from the original quasicrystalline structure. 
		Each site $L_n$ pairs with  site $R_n$ for $n=1,\dots,N_b/2$. 
		By vertically flipping the right-side boundary sites without changing their associated hoppings, 
		the deformed configuration can be viewed as a large-scale unit cell, 
		where the original boundary hoppings now serve as inter-cell hoppings. 
		(c) One-dimensional periodic chain constructed by arranging these unit cells horizontally.
	}
	\label{Fig_reply_chain}
\end{figure}

In fact, the proposed momentum-space Hamiltonian corresponds to a one-dimensional periodic chain 
in which the entire quasicrystalline structure is regarded as a large-scale unit cell. 
As illustrated in Fig.~\ref{Fig_reply_chain}(a), each left-side boundary site $L_n$ pairs with site $R_n$ for $n=1,\dots,N_b/2$, 
where $N_b$ is the number of boundary sites. Note that the right-side boundary sites 
can be vertically flipped without changing their associated hoppings [see Fig.~\ref{Fig_reply_chain}(b)]. 
The resulting deformed configuration can be viewed as a large-scale unit cell, 
with the original boundary hoppings now serving as inter-cell hoppings. 
By arranging these unit cells horizontally [see Fig.~\ref{Fig_reply_chain}(c)], 
we obtain a one-dimensional periodic chain, and the proposed $H(k)$ is precisely the momentum-space 
Hamiltonian of this new chain under periodic boundary conditions. 
Due to the large spatial extension of the unit cell, nontrivial edge states of the chain under open boundary conditions 
are deeply localized within a single unit cell at each terminal. In other words, 
the nontrivial edge states of this chain essentially correspond to those of a single large-scale unit cell, 
i.e., the entire quasicrystalline structure.

\section*{Appendix F: Periodic approximant}
\setcounter{equation}{0} \setcounter{figure}{0} \setcounter{table}{0}
\renewcommand{\theequation}{F\arabic{equation}} \renewcommand{\thefigure}{F\arabic{figure}}
\renewcommand{\theHfigure}{F\arabic{figure}}
\renewcommand{\bibnumfmt}[1]{[#1]} \renewcommand{\citenumfont}[1]{#1}

In this Appendix, we will discuss the periodic approximant of our proposed octagonal quasicrystalline structure 
and show that the eight-fold rotational symmetry plays a critical role in supporting the eight corner modes.

\begin{figure*}[htp]
	\includegraphics[width=0.9\linewidth]{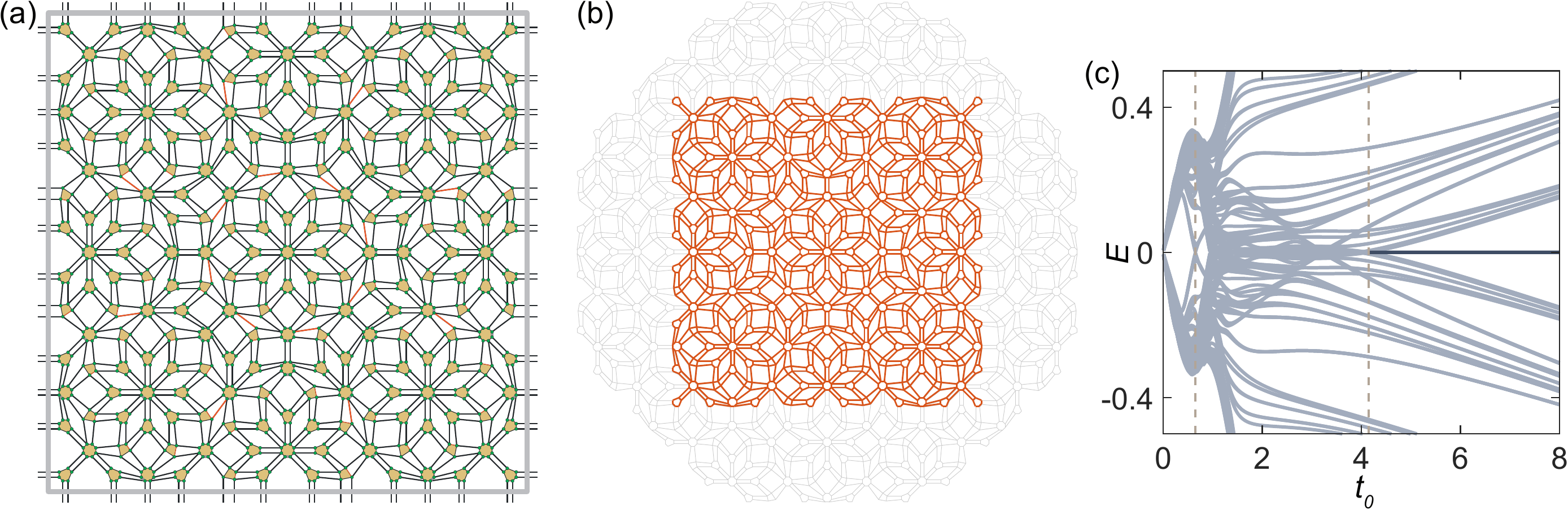}
	\caption{Geometric structure of periodic approximant for octagonal quasicrystalline structure. 
		(a) Illustration of the tight-binding model on the periodic approximant. 
		Red and grey lines represent hoppings with reversed sign and enhanced strength, respectively. 
		Bold grey square denotes the periodic boundary of the entire structure 
		and the total number of lattice sites is 1336. 
		(b) Comparison between the periodic approximant (red lines) 
		and our proposed octagonal quasicrystalline structure (grey lines). 
		(c) Energy spectrum of the tight-binding Hamiltonian 
		shown in (a) under open boundary conditions with respect to $t_0$. 
		The system reveals a trivial regime with two zero-energy modes for $t_0>4.15$, 
		while no gapless corner and edge modes exist in the regime for $t_0<0.65$. 
		Here, all system parameters are the same as those in Fig.~\ref{Fig1}.
	} 
	\label{Fig_reply_approximant}
\end{figure*}

Following the method proposed in Ref.~\cite{jagannathan2024properties}, 
we construct an approximant of the AB tiling octagonal quasicrystal. 
Figure~\ref{Fig_reply_approximant}(a) displays the tight-binding model, 
where the bold grey square represents the periodic boundary. 
This approximant, which includes cell structures, 
is generated using the method described in Appendix B and forms a part of the original quasicrystalline structure, 
as shown in Fig.~\ref{Fig_reply_approximant}(b). 
This model respects the four-fold ($C_4$) and two-fold ($C_2$) rotational symmetries. 
We numerically calculate its eigenenergies and eigenstates, finding that for $t_0<0.65$, 
the system is gapped without gapless edge and corner modes [see Fig.~\ref{Fig_reply_approximant}(c)].

We then obtain a momentum-space Hamiltonian $H_s (k_x,k_y )$ by applying conventional twisted boundary conditions. 
At the high-symmetry points $\Gamma=(0,0)$ and $M=(\pi,\pi)$ in momentum space, 
$H_s$ commutes with the four-fold rotational symmetry operator $U_{C_4}$. 
We thus define the topological invariant as $\chi_{4,p}=N_{4,M}^{(p)}-N_{4,\Gamma}^{(p)}$ for $p=0,\dots,3$, 
where $N_{n,\bm{k}}^{(p)}$ is the number of occupied states below zero energy 
in the eigenspace of $C_n$ symmetry with eigenvalue $\omega_{n,p}=e^{i2\pi p/n}$ at momentum $\bm{k}$. 
Additionally, due to the two-fold ($C_2$) rotational symmetry, 
we define other topological invariants as $\chi_{2,p}=N_{2,X}^{(p)}-N_{2,\Gamma}^{(p)}$ 
for $p=0$ and $1$, where $X=(\pi,0)$. We find that all $\chi_{2,p}$ and $\chi_{4,p}$ vanish for all calculated $t_0$, 
which is consistent with the fact that corner modes are absent. 
These results suggest the critical role of eight-fold rotational symmetry in supporting corner modes in our model.

In addition, we observe two zero-energy states when $t_0>4.15$. Their presence
does not contradict the proof in Appendix E, which establishes that a pair of zero-energy
states must exist in a system with eight-fold rotational symmetry, time-reversal symmetry, 
and chiral symmetry, provided that the $1/8$ sector contains an odd number of sites. 
Crucially, that proof does not specify the locations of these states. 
Accordingly, two exact zero-energy states exist for all $t_0$ as shown in Fig.~1(b) in the main text; 
however, their positions depend on the value of $t_0$. 
In contrast, in Fig.~\ref{Fig_reply_approximant}(c),
zero-energy states appear only for sufficiently large $t_0$, indicating that their existence is
not protected by these symmetries---indeed, the approximant does not respect 
eight-fold rotational symmetry.
Their emergence is readily understood as a remnant of the zero-energy states 
localized at the center of the original quasicrystalline lattice in the large-$t_0$ regime.

\section*{Appendix G: Experimental Details}
\setcounter{equation}{0} \setcounter{figure}{0} \setcounter{table}{0}
\renewcommand{\theequation}{G\arabic{equation}} \renewcommand{\thefigure}{G\arabic{figure}}
\renewcommand{\theHfigure}{G\arabic{figure}}
\renewcommand{\bibnumfmt}[1]{[#1]} \renewcommand{\citenumfont}[1]{#1}

In this Appendix, we will discuss the specific design of cavity couplings for realizing the tight-binding Hamiltonian.

\begin{figure}[htp]
	\includegraphics[width=0.9\linewidth]{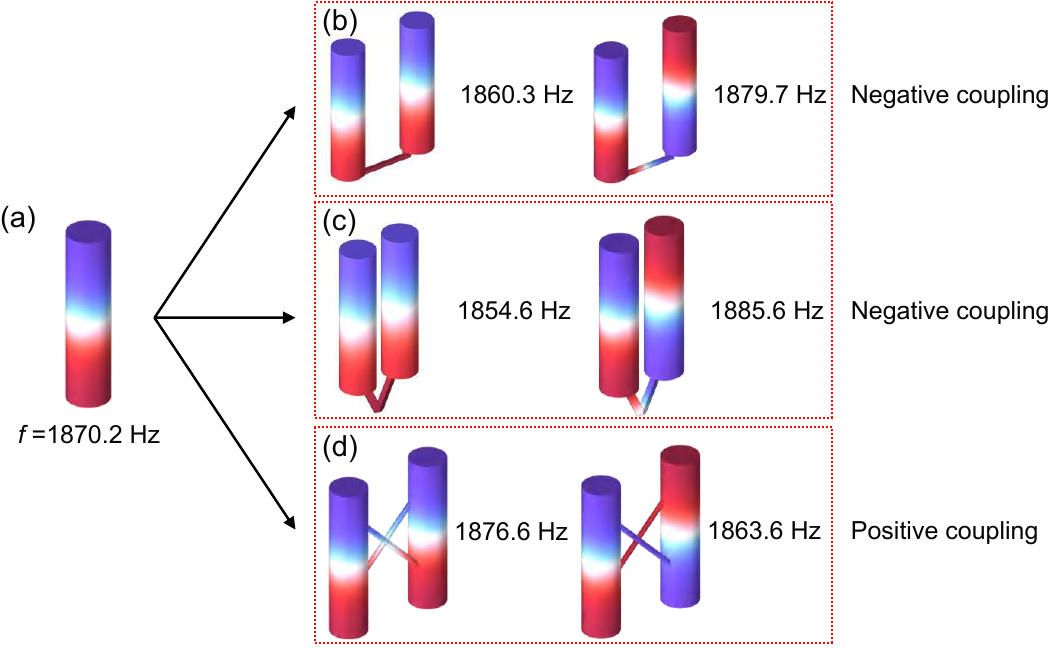}
	\caption{{Illustration of the positive and negative couplings.} 
		(a) Acoustic dipole model in a single cavity at $f=1870.2$ Hz. 
		(b)--(c) Negative coupling realized by (b) straight and (c) bent tubes. 
		(d) Positive coupling realized by a pair of cross-linked tubes.} 
	\label{FigSM_exp_couple}
\end{figure}

\begin{figure}[htp]
	\includegraphics[width=0.95\linewidth]{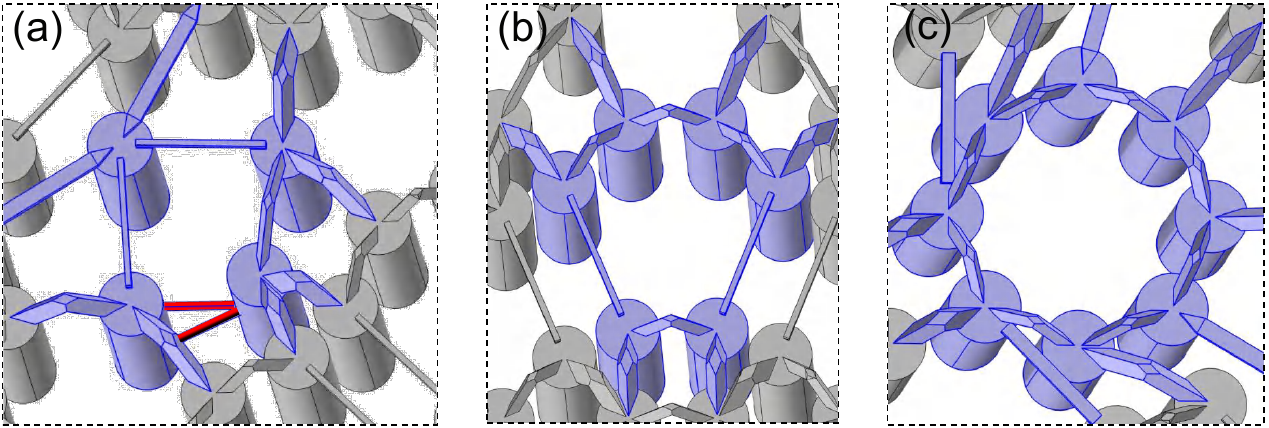}
	\caption{{Cell structures in experiments.} 
		The light purple structures describe the 
		four-site cell in (a), six-site cell in (b), and eight-site cell in (c) 
		and their couplings in the acoustic quasicrystal. 
		The cross sections of straight tubes are rectangular, whereas those of bent tubes are triangular. 
		Red tubes denote the sign-reversal coupling.} 
	\label{FigSM_exp_cell}
\end{figure}

For the acoustic dipole mode of the cylindrical cavity shown in Fig.~\ref{FigSM_exp_couple}(a), 
the straight connection with thin tubes results in the splitting of the resonant frequency 
corresponding to in-phase (IP) and out-of-phase (OP) distributions. 
A negative coupling indicates that the IP distribution has a lower frequency~\cite{ni2020demonstration,Qi2020PRL} 
[see Fig.~\ref{FigSM_exp_couple}(b)]. We notice that for the octagonal quasicrystalline structure 
shown in Fig.~\ref{Fig1}(a), the straight tube shifts the resonant frequency when two cavities are in close proximity. 
Thus, as shown in Fig.~\ref{FigSM_exp_couple}(c), we bend the tube such that the center frequency is fixed near $f=1870$ Hz. 
In addition, to realize sign-reversal couplings, two cavities are connected by a pair of cross-linked tubes, 
thereby reversing the frequency order of the IP and OP distributions 
[see Fig.~\ref{FigSM_exp_couple}(d)]. This configuration thus corresponds to the sign-reversal modification. 
Based on the above schemes, we design and fabricate the acoustic quasicrystal [see Fig.~\ref{Fig2}(a)], 
and the specific structures of different cells are displayed in Fig.~\ref{FigSM_exp_cell}.

\begin{figure}[htp]
	\includegraphics[width=0.95\linewidth]{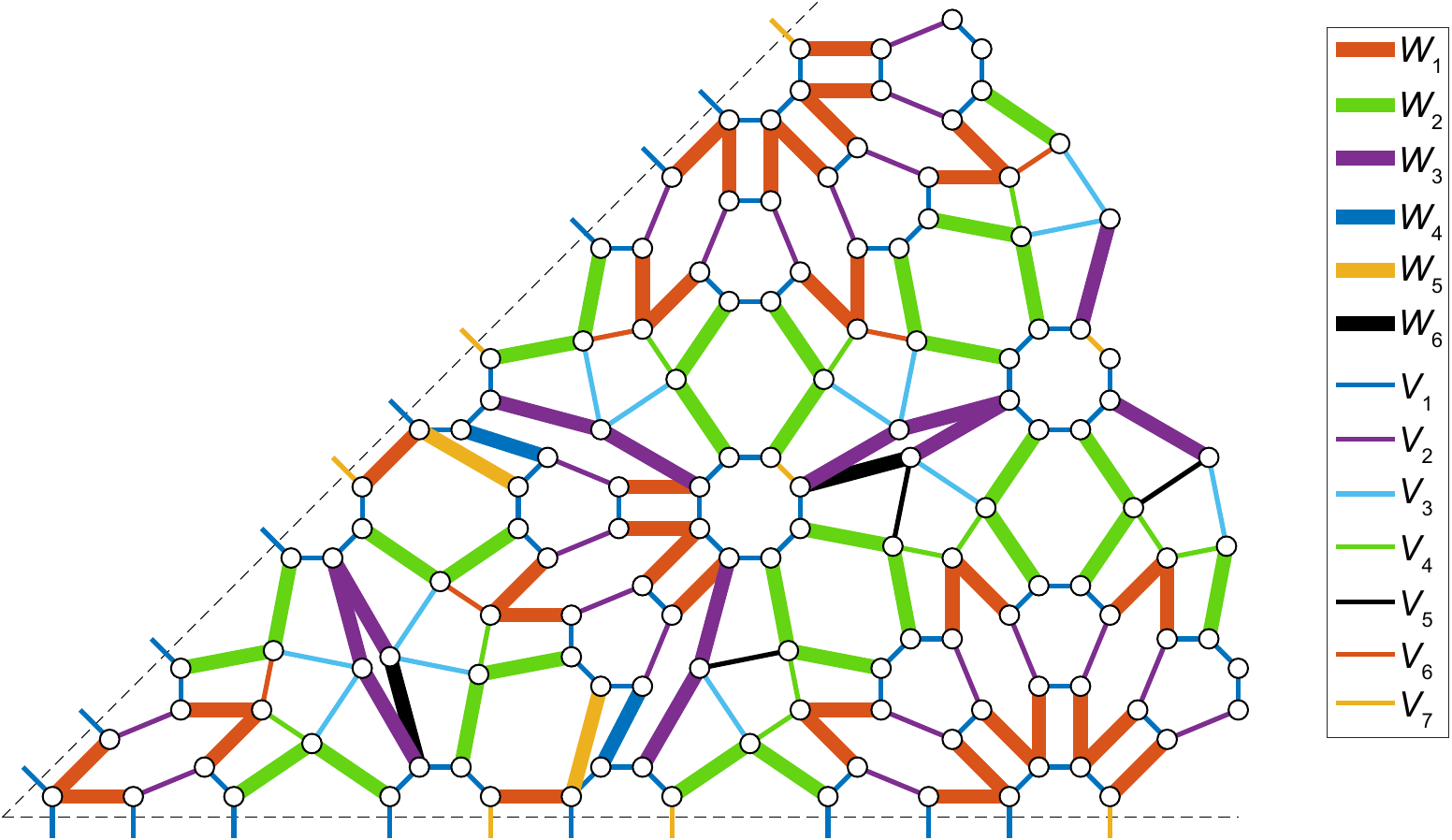}
	\caption{{All types of couplings in the topological acoustic quasicrystal.} 
		This figure illustrates the 1/8 sector of the acoustic quasicrystal. 
		Colored lines denote different types of couplings, 
		and $W$ and $V$ represent inter-cell and intra-cell couplings, respectively.} 
	\label{FigSM_exp_TB}
\end{figure}

\begin{figure}[htp]
	\includegraphics[width=0.7\linewidth]{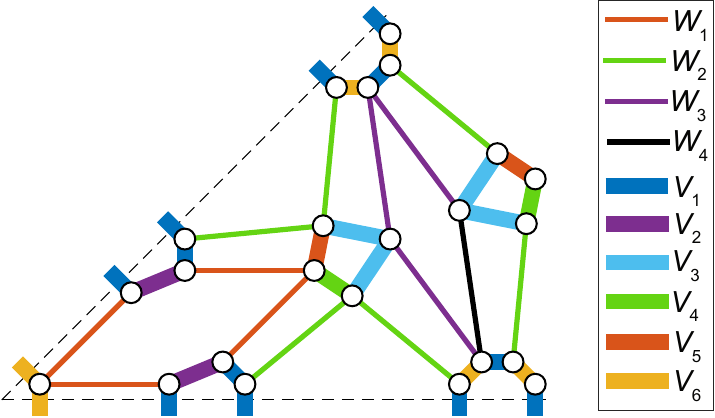}
	\caption{{All types of couplings in the trivial acoustic quasicrystal.} 
		This figure illustrates the 1/8 sector of the trivial acoustic quasicrystal. 
		Colored lines denote different types of couplings, and $W$ and $V$ represent inter-cell and intra-cell couplings, respectively.} 
	\label{FigSM_exp_TB_trivial}
\end{figure}

\begin{figure*}[htp]
	\includegraphics[width=1\linewidth]{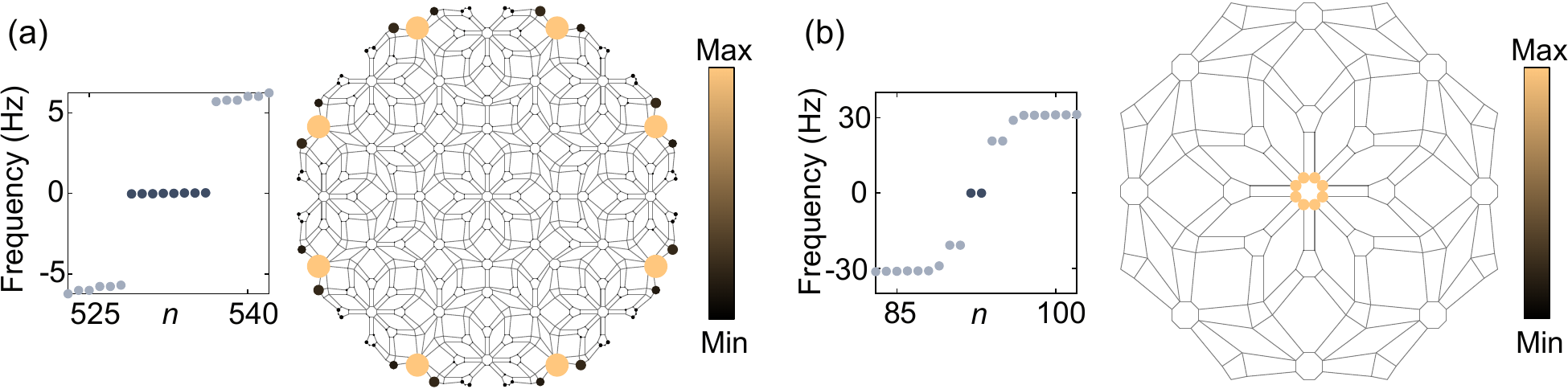}
	\caption{{Simulation results for the tight-binding Hamiltonians constructed based on the data listed 
			in Table~\ref{table_nontrivial} and Table~\ref{table_trivial}.} 
		Eigenfrequency spectrum versus state index $n$ and spatial distribution of the local DOS at zero frequency
		for (a) nontrivial model and for (b) trivial model. } 
	\label{FigSM_sim_TB}
\end{figure*}

To verify the corresponding tight-binding Hamiltonians of the topological and trivial acoustic quasicrystals, we simulate all possible couplings between cavities (colored lines in Figs.~\ref{FigSM_exp_TB} and \ref{FigSM_exp_TB_trivial}), with the resulting strengths listed in Tables~\ref{table_nontrivial} and \ref{table_trivial}. We construct the tight-binding Hamiltonians from the data and present the resulting eigenfrequencies and local DOS at zero frequency in Fig.~\ref{FigSM_sim_TB}, which agree well with the experimental observations.

\begin{figure}[htbp]
	\includegraphics[width=0.45\linewidth]{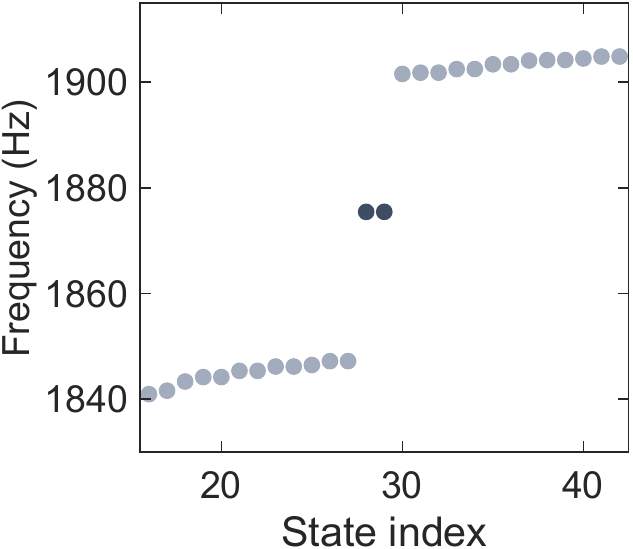}
	\caption{Simulated eigenfrequencies of the topologically trivial acoustic quasicrystal 
		by slightly enhancing the intra-cell coupling strength within the central cell. 
		The middle pair of zero-energy modes (black circles) remains in the gap.
	} 
	\label{Fig_reply_middle_zero_mode}
\end{figure}

In addition, for the trivial acoustic quasicrystal, only the middle two degenerate states 
in Fig.~3(b) in the main text are zero-energy states. 
The other two pairs merge into the continuum spectrum when the intra-cell coupling strength is increased, 
as shown in Fig.~\ref{Fig_reply_middle_zero_mode}.

\begin{table*}[htbp]
	\centering % 
	\begin{tabular}{c|cccc}
		\toprule
		Connecting tube & In-phase frequency (Hz) & Out-of-phase frequency (Hz) & Center frequency (Hz) & Coupling strength (Hz) \\
		\hline 
		$W_1$   & 1827.3 & 1913.3 & 1870.3  &-43 \\
		$W_2$   & 1845   & 1895.3 & 1870.15 &-25.15 \\
		$W_3$   & 1846.2 & 1894.2 & 1870.2  &-24 \\
		$W_4$   & 1837.6 & 1902.9 & 1870.25 &-32.65 \\
		$W_5$   & 1842.1 & 1897.1 & 1869.6  &-27.5 \\
		$W_6$   & 1891.1 & 1848.8 & 1869.95 &21.15 \\
		$V_1$   & 1860.9 & 1880.1 & 1870.5  &-9.6 \\
		$V_2$   & 1860.7 & 1880   & 1870.35 &-9.65 \\
		$V_3$   & 1860.3 & 1879.9 & 1870.1  &-9.8 \\
		$V_4$   & 1863.9 & 1876.6 & 1870.25 &-6.35 \\
		$V_5$   & 1876.6 & 1863.6 & 1870.1  &6.5 \\
		$V_6$   & 1879 & 1861.2   & 1870.1  &8.9 \\
		$V_7$   & 1860.6 & 1880.7 & 1870.65 &-10.05 \\
		\bottomrule
	\end{tabular}
	\caption{The specific coupling strengths in the topological acoustic quasicrystal.}
	\label{table_nontrivial}
\end{table*}

\begin{table*}[htbp]
	\centering % 
	\begin{tabular}{c|cccc}
		\toprule
		Connecting tube & In-phase frequency (Hz) & Out-of-phase frequency (Hz) & Center frequency (Hz) & Coupling strength (Hz) \\
		\hline 
		$W_1$   & 1864.7 & 1875.9 & 1870.3  &-5.6 \\
		$W_2$   & 1865.8 & 1875   & 1870.4  &-4.6 \\
		$W_3$   & 1862   & 1878.3 & 1870.15 &-8.15 \\
		$W_4$   & 1872.6 & 1867.7 & 1870.15 &2.45 \\ 
		$V_1$   & 1816.2 & 1924   & 1870.1  &-53.9 \\
		$V_2$   & 1846.2 & 1893.7 & 1869.95 &-23.75 \\
		$V_3$   & 1845.4 & 1893.7 & 1869.55 &-24.15 \\
		$V_4$   & 1839.1 & 1901.7 & 1870.4  &-31.3 \\
		$V_5$   & 1903.3 & 1836.3 & 1869.8  &33.5 \\
		$V_6$   & 1853.9 & 1885.1 & 1869.5  &-15.6 \\
		\bottomrule
	\end{tabular}
	\caption{The specific coupling strengths in the trivial acoustic quasicrystal.}
	\label{table_trivial}
\end{table*}

\section*{Appendix H: Comparison of the Fourier spectrum of the measured acoustic pressure field distribution with that of the quasicrystalline lattice distribution}
\setcounter{equation}{0} \setcounter{figure}{0} \setcounter{table}{0}
\renewcommand{\theequation}{H\arabic{equation}} \renewcommand{\thefigure}{H\arabic{figure}}
\renewcommand{\theHfigure}{H\arabic{figure}}
\renewcommand{\bibnumfmt}[1]{[#1]} \renewcommand{\citenumfont}[1]{#1}

\begin{figure*}[htbp]
	\includegraphics[width=0.8\linewidth]{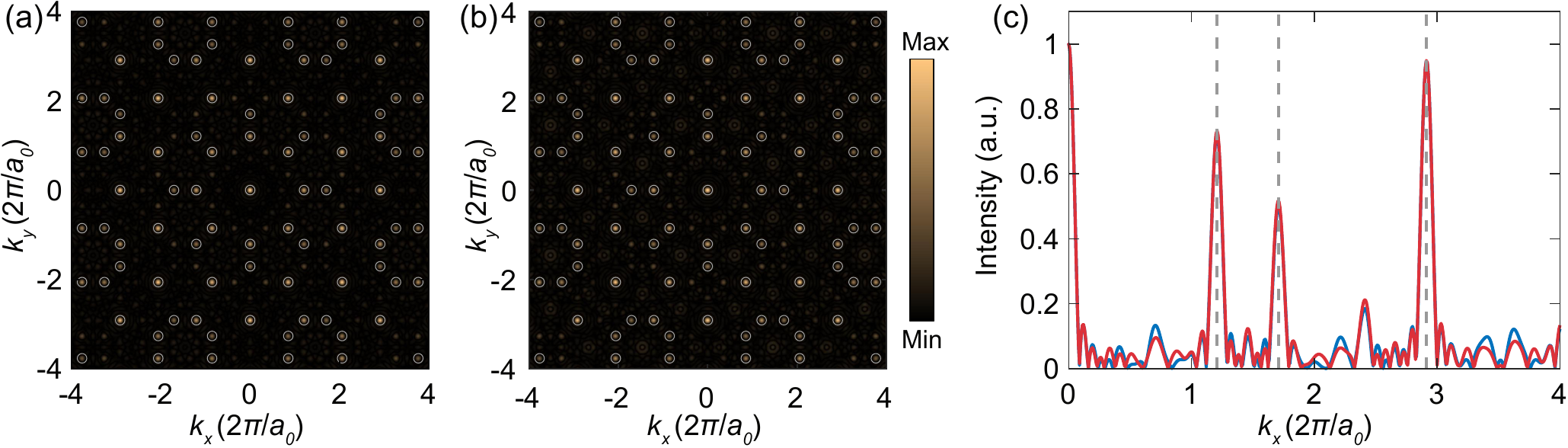}
	\caption{{Comparison of the Fourier spectrum.} 
		(a)--(b) Fourier spectra of the spatial distribution of the octagonal quasicrystalline lattice (a) 
		and that of the measured pressure field (b) in the $k_x$--$k_y$ space. 
		Small circles highlight the principle peaks. 
		(c) Plot of (a) as a blue line and (b) as a red line along $k_x$ at $k_y=0$.
		Dashed lines show the peak positions with $k_x/(2\pi/a_0)=(\sqrt{2}+1)/2,\ (\sqrt{2}+2)/2$, and $(2\sqrt{2}+3)/2$.} 
	\label{FigSM_AB_FT}
\end{figure*}

The Fourier spectrum of the measured acoustic pressure field distribution [Fig.~\ref{FigSM_AB_FT}(b)]
exhibits a distinct pattern of peaks, 
which aligns closely with the Fourier spectrum 
of the quasicrystalline lattice, defined as $F_{\mathrm{oct} }(\bm{k})=\sum_{v}e^{-i\bm{k}\cdot\bm{r}_v}$~\cite{Kuo1987PRL}
[see Fig.~\ref{FigSM_AB_FT}(a)].
This agreement is quantified in Fig.~\ref{FigSM_AB_FT}(c),
which compares the spectra along the line $k_y=0$. 
The principle peaks appear 
at $k_x/(2\pi/a_0)=(\sqrt{2}+1)/2,\ (\sqrt{2}+2)/2$, and $(2\sqrt{2}+3)/2$.

\begin{figure*}[htbp]
	\includegraphics[width=0.8\linewidth]{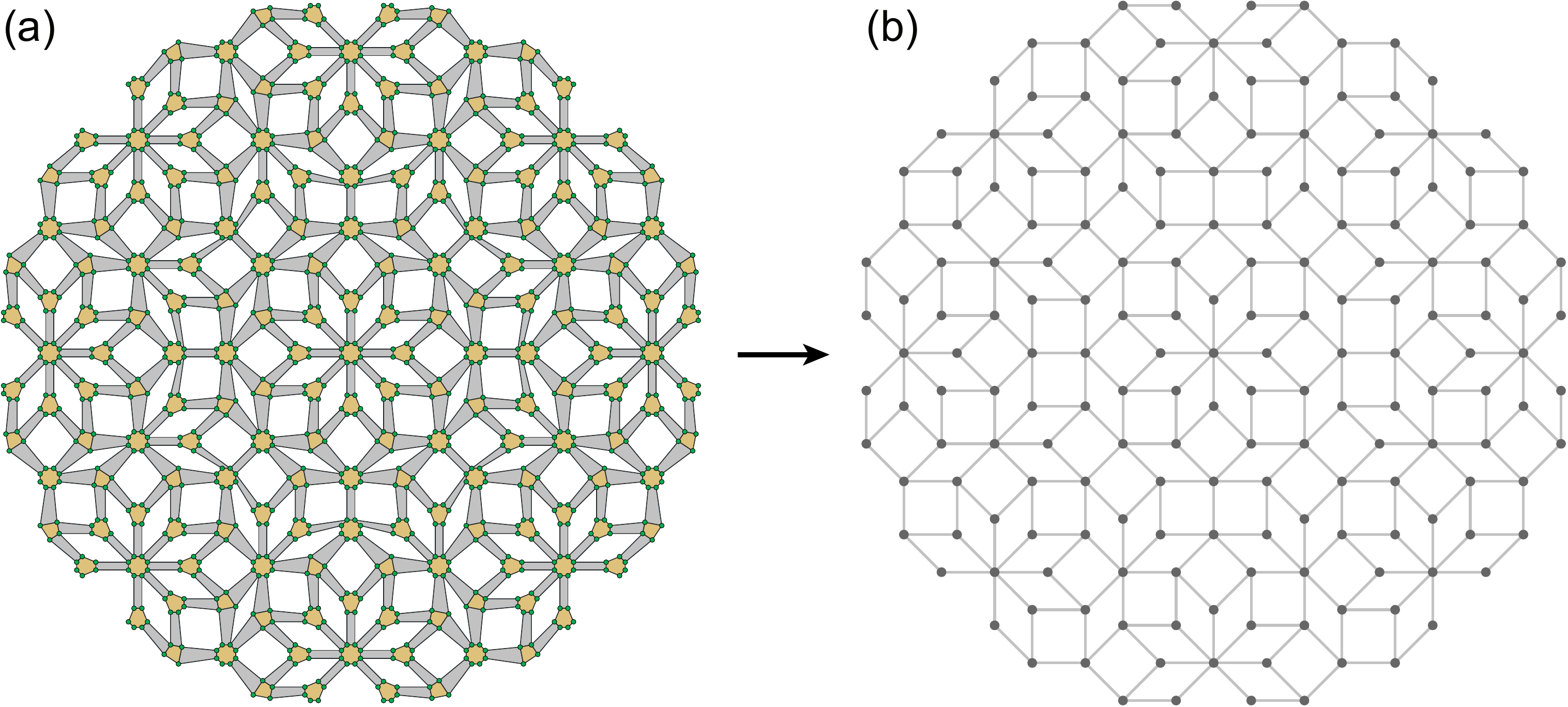}
	\caption{Visualization of the quasicrystalline nature. With all cells contracted to points, 
		the gray regions related to inter-cell bonds reduce to lines, 
		and the resulting structure becomes the AB tiling octagonal quasicrystal.
	}
	\label{Fig_reply_cell_to_vertex}
\end{figure*}

In addition, quasicrystalline nature can be visually confirmed via cell deformation. 
As shown in Fig.~\ref{Fig_reply_cell_to_vertex}(a), we fill the bonds between two cells with gray. 
When all cells contract to points, the gray region reduces to a line, 
transforming the structure to the AB tiling octagonal quasicrystal [see Fig.~\ref{Fig_reply_cell_to_vertex}(b)]. 
This quasicrystal consists solely of rhombuses and squares, providing visual confirmation of its quasicrystalline nature.

\section*{Appendix I: Effects of acoustic loss in the acoustic quasicrystals}
\setcounter{equation}{0} \setcounter{figure}{0} \setcounter{table}{0}
\renewcommand{\theequation}{I\arabic{equation}} \renewcommand{\thefigure}{I\arabic{figure}}
\renewcommand{\theHfigure}{I\arabic{figure}}
\renewcommand{\bibnumfmt}[1]{[#1]} \renewcommand{\citenumfont}[1]{#1}

\begin{figure}[htp]
	\includegraphics[width=1\linewidth]{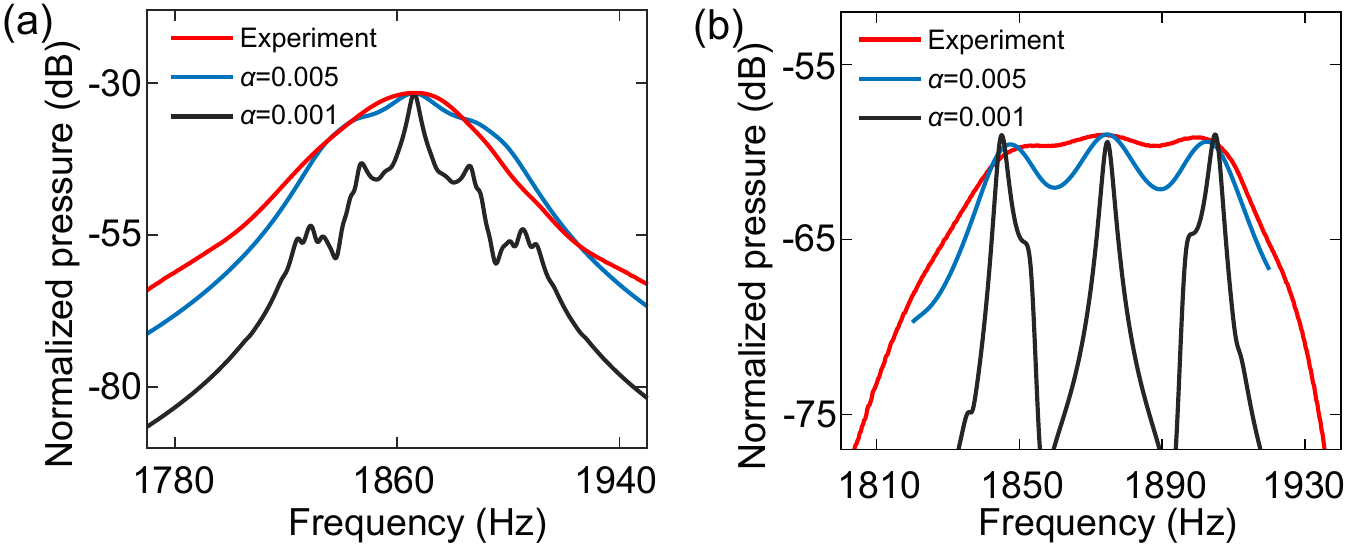}
	\caption{{Measured and simulated response spectra.} 
		Measured (red line) and simulated response spectra for different loss 
		coefficients $\alpha$ in (a) topological acoustic quasicrystal
		and (b) trivial acoustic quasicrystal.} 
	\label{FigSM_exp_loss}
\end{figure}

In the main text, we show that incorporating acoustic loss explains the significant broadening of 
the measured response spectrum. Here, we provide the numerically simulated response spectra with 
a reduced loss coefficient 
$\alpha$ in Fig.~\ref{FigSM_exp_loss}. We see that as $\alpha$ decreases, the  
response spectra display distinct peaks at the frequencies of the corner modes and 
localized modes at the center. This indicates that reducing acoustic loss can more 
clearly reveal the zero-energy modes.

\section*{Appendix J: Details of experimental realization of the trivial quasicrystal}
\setcounter{equation}{0} \setcounter{figure}{0} \setcounter{table}{0}
\renewcommand{\theequation}{J\arabic{equation}} \renewcommand{\thefigure}{J\arabic{figure}}
\renewcommand{\theHfigure}{J\arabic{figure}}
\renewcommand{\bibnumfmt}[1]{[#1]} \renewcommand{\citenumfont}[1]{#1}

\begin{figure*}[htp]
	\includegraphics[width=0.8\linewidth]{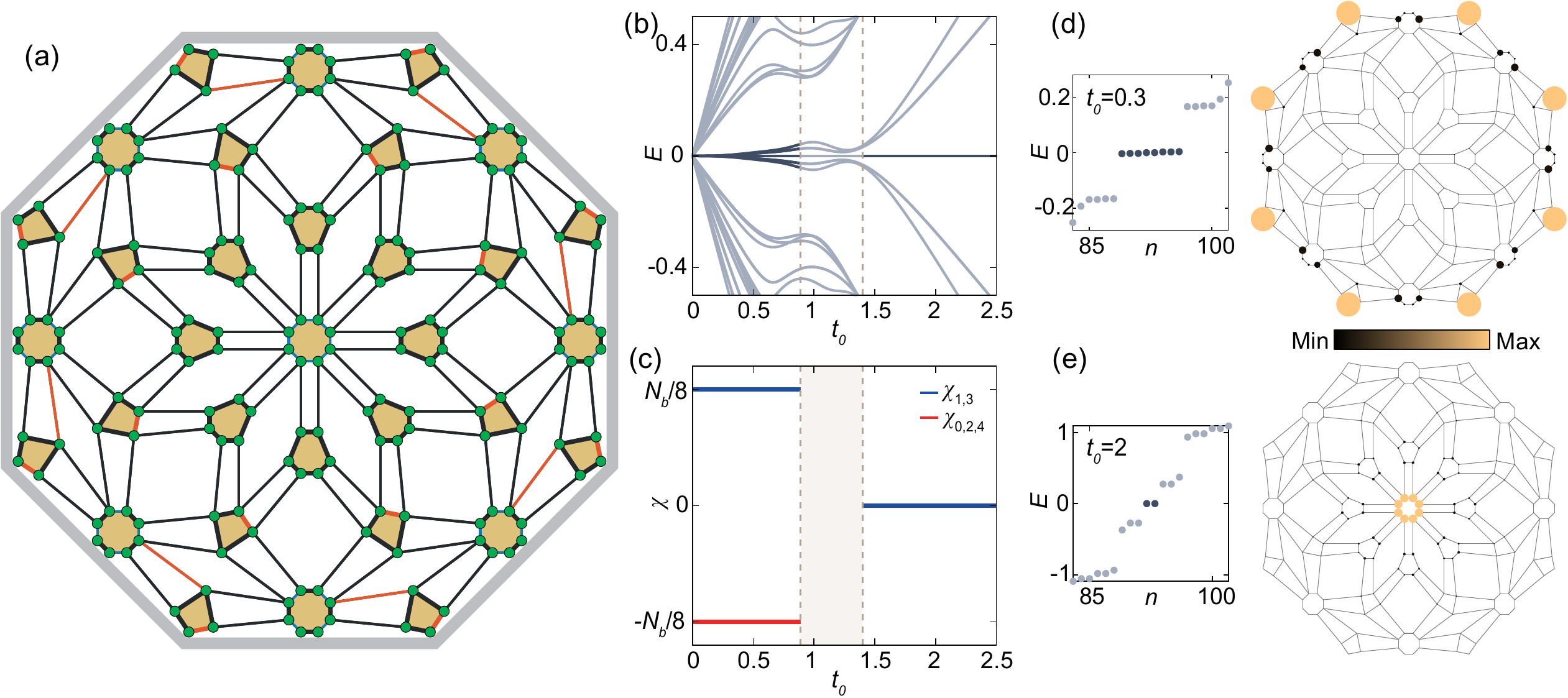}
	\caption{{Quasicrystalline structure for observing the trivial phase.} 
		(a) Illustration of the tight-binding model on the octagonal quasicrystalline lattice. 
		Compared to Fig.~\ref{Fig1}(a), we adjust certain intra-cell and inter-cell hoppings
		by altering the value of $\xi$. Specifically, the blue bonds denote $\xi=0.3$, 
		while the red ones denote $\xi=-1$. The total number of sites is $N=184$. 
		(b)	Energy spectrum of the tight-binding Hamiltonian shown in (a) with respect 
		to $t_0$. Similar to the case described in the main text, the system reveals a topological 
		regime with eight zero-energy corner modes for $0<t_0<0.89$ and a trivial regime 
		with two zero-energy modes for $t_0>1.4$. 
		(c) Topological invariants $\chi_p$ as a function of $t_0$ with $p=0,\dots,4$. 
		In the topological regime, $\chi_p=\pm N_b/8$ with $N_b=24$, whereas in the trivial regime, 
		$\chi_p=0$. 
		(d)--(e) Energy spectrum versus the state index $n$ and spatial distribution of the local 
		DOS at zero energy at (d) $t_0=0.3$ and at (e) $t_0=2$. 
		The size of the circles represent the magnitude of the local DOS, complementing 
		the color representation.
	}
	\label{FigSM_C8_adjust_1_chi}
\end{figure*}

To fabricate a trivial acoustic quasicrystal for observation, we select a 
different set of coefficients $\xi_{vm,wn}$ in Eq.~(\ref{H_TB}) (see  
the caption of Fig.~\ref{FigSM_C8_adjust_1_chi} for details).
The resulting phase diagram, shown in Fig.~\ref{FigSM_C8_adjust_1_chi}(b) and (c),
is similar to that in Fig.~\ref{Fig1}.

\section*{Appendix K: Asymmetric effects}
\setcounter{equation}{0} \setcounter{figure}{0} \setcounter{table}{0}
\renewcommand{\theequation}{K\arabic{equation}} \renewcommand{\thefigure}{K\arabic{figure}}
\renewcommand{\theHfigure}{K\arabic{figure}}
\renewcommand{\bibnumfmt}[1]{[#1]} \renewcommand{\citenumfont}[1]{#1}

In our measurement shown in Fig.~3(a), we use only a single loudspeaker, which 
breaks the eight-fold rotational symmetry. In this Appendix, we will discuss this effect.

\begin{figure*}[htp]
	\includegraphics[width=0.8\linewidth]{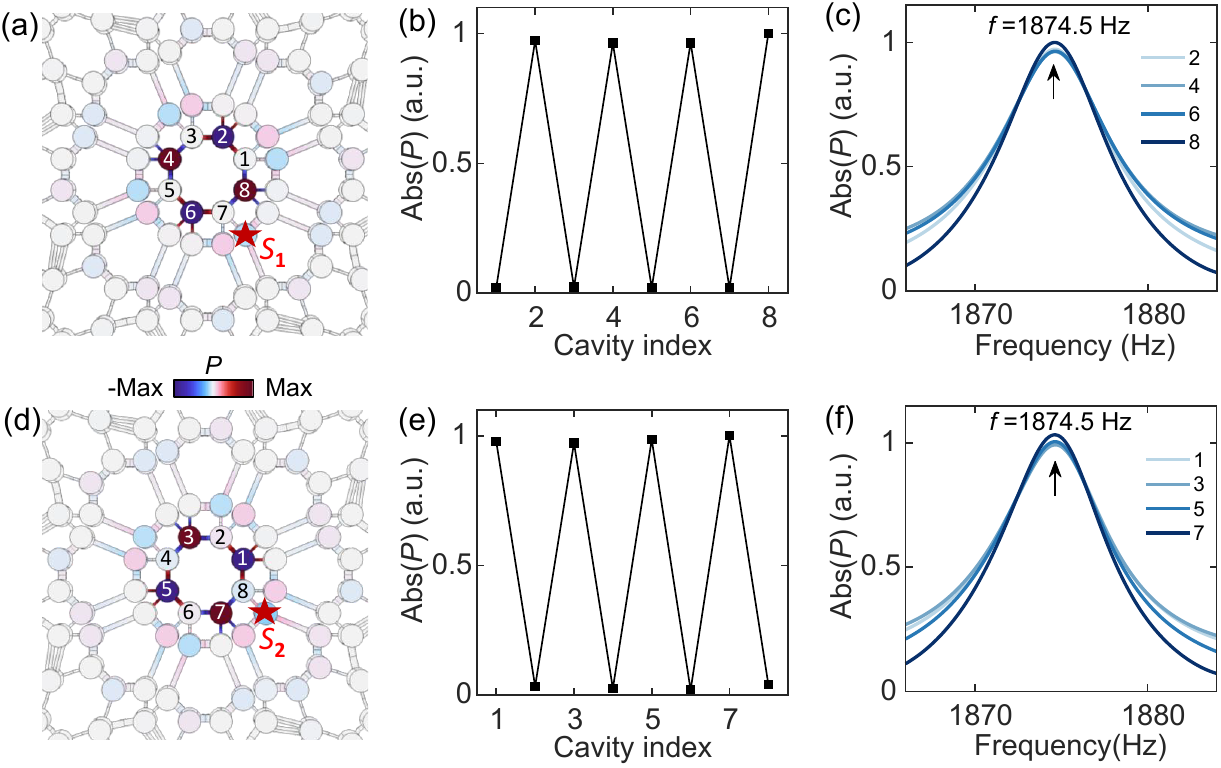}
	\caption{Simulated results for the topologically trivial acoustic quasicrystal with a single source. 
		(a) Simulated acoustic pressure field distribution (without loss) by placing the source 
		at cavity $S_1$ near the center at the frequency of 1874.5 Hz. 
		The pressure field is mainly localized in the central even cavities. 
		(b)--(c) Extracted acoustic pressure data and response spectra from the central cavities in (a). 
		(d)--(f) The same as (a)--(c) but with the source placed at $S_2$, 
		in which the pressure field is mainly localized in the odd cavities.
	} 
	\label{Fig_reply_simulation_central_cell}
\end{figure*}

We simulate the excited acoustic pressure field distribution at cavities labelled by $1,2,\dots,8$ 
by placing a source at $S_1$ [see Fig.~\ref{Fig_reply_simulation_central_cell}(a)] or 
$S_2$ [see Fig.~\ref{Fig_reply_simulation_central_cell}(d)] near the center at the frequency of 1874.5 Hz. 
The corresponding acoustic pressure data at the central eight cavities are presented in 
Figs.~\ref{Fig_reply_simulation_central_cell}(b) and (e) for sources at $S_1$ and $S_2$, 
respectively. We see that for the former case, 
the pressure field exists only in the even cavities, while for the latter, it is nonzero only in the odd cavities. 
The near-uniform magnitudes observed indicate a negligible asymmetric effect 
(an explanation for the excitation of only even or odd cavities follows). Additionally, 
simulated response spectra as a function of frequency at these central cavities 
[see Figs.~\ref{Fig_reply_simulation_central_cell}(c) and (f)] 
reveal that their peak frequencies align with those of the zero-energy modes, 
with only minor magnitude fluctuations. This further confirms the minimal impact of asymmetry.

\begin{figure}[htp]
	\includegraphics[width=1\linewidth]{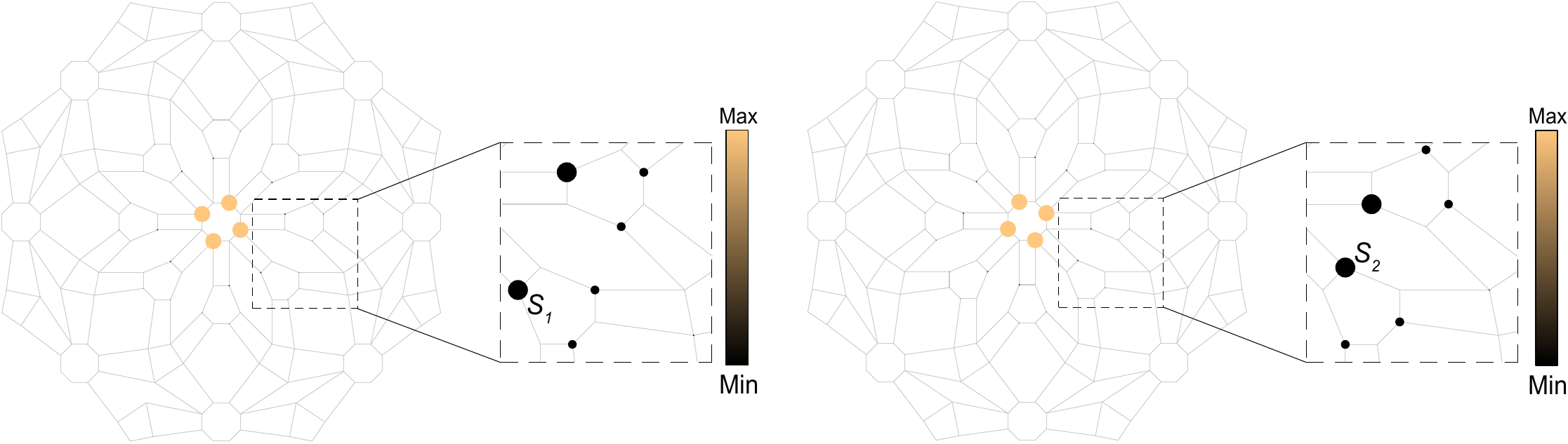}
	\caption{Spatial distributions of the two theoretical degenerate zero-energy modes 
		in the tight-binding Hamiltonian for the topologically trivial acoustic quasicrystal. 
		The size of the circles represents the magnitude of their amplitudes. 
		Each inset provides a zoomed-in view of an area near the center. 
		Here, the tight-binding Hamiltonian is constructed based on the data listed in Table~\ref{table_trivial}.
	}
	\label{Fig_reply_simulation_trivial}
\end{figure}

We now explain why the excitations only exist in even or odd cavities by examining the 
two zero-energy eigenstates $\psi_1$ and $\psi_2$, localized at the center 
(see Fig.~\ref{Fig_reply_simulation_trivial} for their spatial distributions). 
In the figure, the circle sizes correspond to the amplitude magnitudes of each state with insets 
providing a zoomed-in view of an area near the center. 
We see that the state $\psi_1$ has nonzero amplitude in the central even cavities, 
whereas $\psi_2$ is nonzero in the central odd cavities. Crucially, 
$\psi_1$ exhibits a small but finite amplitude at the position of source $S_1$, 
enabling $S_1$ to excite the pressure field solely within the even cavities. 
Similarly, the small amplitude of $\psi_2$ at source $S_2$ allows $S_2$ to excite only the odd cavities.

For the acoustic quasicrystal in a topological phase, 
the simultaneous excitation of all corner cavities with a single loudspeaker 
is impractical owing to substantial inter-corner distances and experimental losses. 
Therefore, we measure the response spectrum for an individual corner 
by introducing excitation in a nearby cavity, as illustrated in Fig.~\ref{Fig2}. 
The measured corner mode frequency appears as the central peak in the response spectrum, 
in agreement with the simulated eigenfrequency for the corner modes of the full structure.

\section*{Appendix L: Effects of structural defects and disorder on topological corner modes}
\setcounter{equation}{0} \setcounter{figure}{0} \setcounter{table}{0}
\renewcommand{\theequation}{L\arabic{equation}} \renewcommand{\thefigure}{L\arabic{figure}}
\renewcommand{\theHfigure}{L\arabic{figure}}
\renewcommand{\bibnumfmt}[1]{[#1]} \renewcommand{\citenumfont}[1]{#1}

In this Appendix, we will examine the robustness of the topological phase against structural defects and coupling disorder. 
We find that the topological phase is stable under weak disorder. Detailed analysis is presented below.

\begin{figure*}[htp]
	\includegraphics[width=0.8\linewidth]{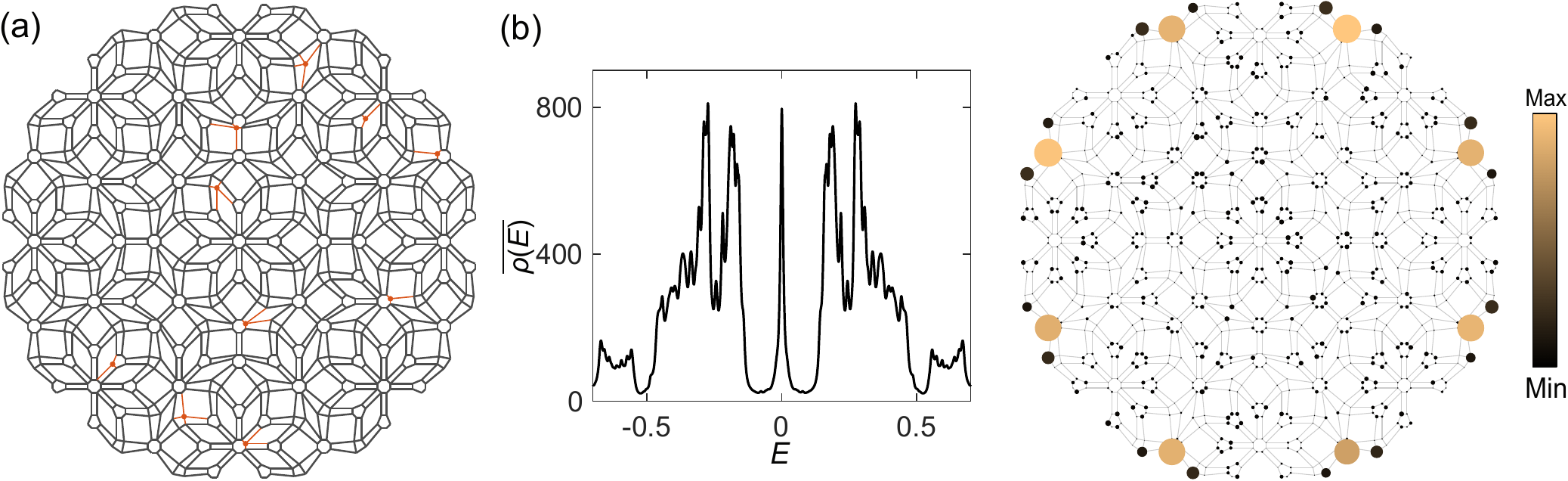}
	\caption{Results for the tight-binding model with structural defects.
		(a) Illustration of the quasicrystalline structure with random structural defects. 
		Red points and lines represent the randomly removed sites and bonds. 
		(b) Spectrum of the averaged density of states $\overline{\rho(E)}$ 
		with $\overline{\rho(E)}=\sum_{\textbf{r}}\rho(E,\textbf{r})$ as a function of $E$ and spatial distribution 
		of the averaged local DOS at zero energy for the quasicrystalline structure with structural defects. 
		The number of removed sites is 10. Here, the system parameters are the same as those 
		in Fig.~\ref{Fig1} with $t_0=0.3$. The number of random configurations is 100.
	}
	\label{Fig_reply_perturbation_defect}
\end{figure*}

We first consider the structural defects. As illustrated in Fig.~\ref{Fig_reply_perturbation_defect}(a), 
we randomly remove 10 lattice sites (red-filled circles) and their associated bonds (red bonds). 
The resulting density of states $\overline{\rho(E)}$, averaged over 100 random configurations, 
is shown in Fig.~\ref{Fig_reply_perturbation_defect}(b). 
The averaged density of states exhibits a sharp peak at zero energy, 
with the corresponding spatial distribution localized at the eight corners. 
This confirms the robustness of the topological corner modes against structural defects in our quasicrystalline system.

\begin{figure}[htp]
	\includegraphics[width=0.8\linewidth]{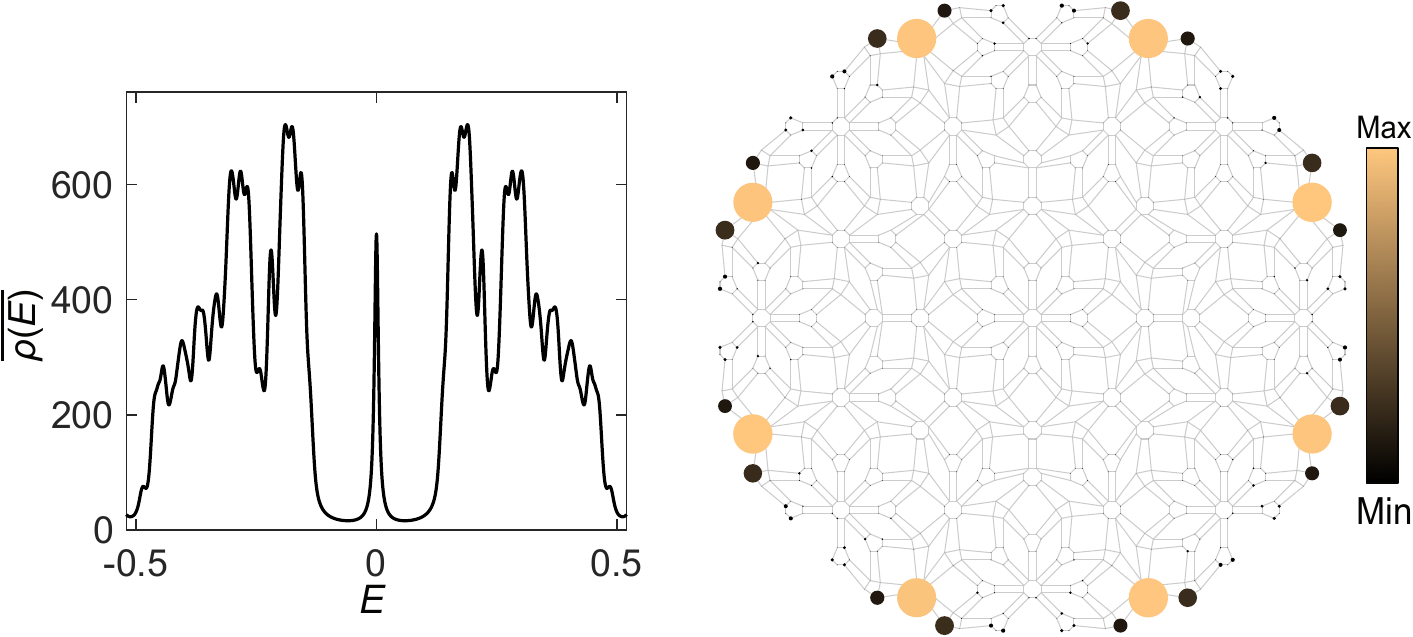}
	\caption{Results for the tight-binding model with disordered coupling strength. 
		Spectrum of the averaged density of states $\overline{\rho(E)}$ as a function of $E$ and 
		spatial distribution of the averaged local DOS at zero energy. 
		Here, we set $W_t=0.1$ and the system parameters are the same as those in Fig.~1 in the main text with $t_0=0.3$. 
		The number of random configurations is 100.
	}
	\label{Fig_reply_perturbation_disorder}
\end{figure}

In addition, we consider coupling disorder. 
We introduce coupling disorder $t_{ij}^{(0)}\delta t_{ij}$ between cavities $i$ and $j$ 
so that the total coupling strength is $t_{ij}^{(d)}=t_{ij}^{(0)}(1+\delta t_{ij})$ with $\delta t_{ij}=W_t u_{ij}$, 
where $t_{ij}^{(0)}$ is the ideal coupling strength, $W_t$ is the disorder strength, 
and $u_{ij}$ is an uncorrelated uniform random variable within $[-1,1]$. In this analysis, 
$W_t$ is set to 0.1 without including structural defects. 
The density of states, averaged over 100 random configurations (see Fig.\ref{Fig_reply_perturbation_disorder}), 
confirms that the topological corner modes remain robust against such disorder.

\section*{Appendix M: Decagonal quasicrystalline structure}
\setcounter{equation}{0} \setcounter{figure}{0} \setcounter{table}{0}
\renewcommand{\theequation}{M\arabic{equation}} \renewcommand{\thefigure}{M\arabic{figure}}
\renewcommand{\theHfigure}{M\arabic{figure}}
\renewcommand{\bibnumfmt}[1]{[#1]} \renewcommand{\citenumfont}[1]{#1}

In this Appendix, we will construct a decagonal quasicrystalline structure and 
demonstrate the presence of ten near zero-energy corner modes in the topological regime.

\begin{figure*}[htp]
	\includegraphics[width=0.95\linewidth]{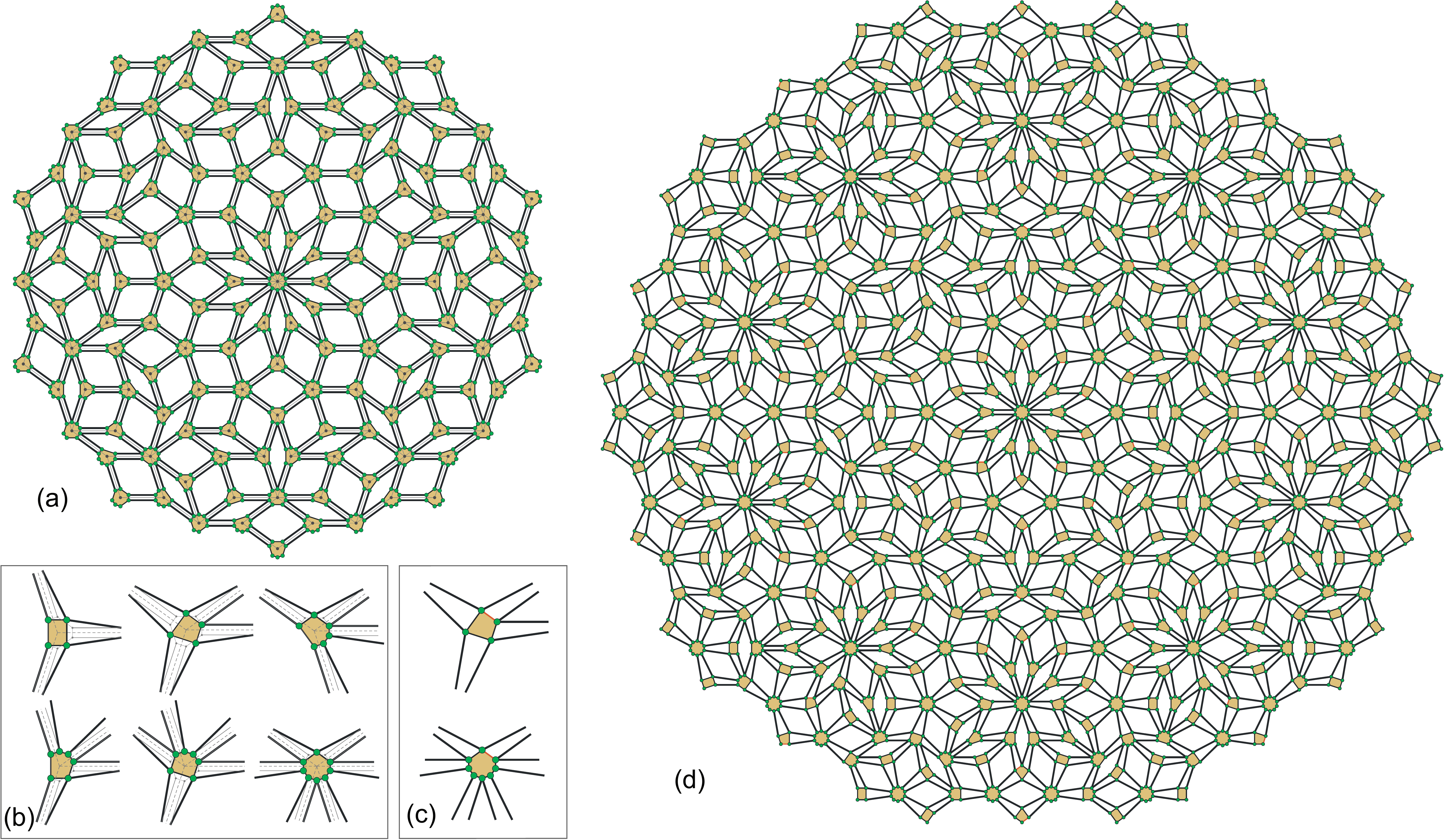}
	\caption{{Construction of a tight-binding model on a decagonal quasicrystalline lattice.}
		(a) The resulting structure obtained by applying the following operations.
		First, we generate the Penrose tiling decagonal quasicrystalline lattice described by 
		grey circles and connecting lines.
		At each lattice vertex, we place a cell containing ten sites, uniformly
		distributed on a small circle centered at the vertex.
		After introducing inter-cell bonds as described, sites lacking inter-cell bonds 
		are removed, and intra-cell bonds are introduced for the remaining sites. 
		(b) We list all types of cell structures that are modified. 
		(c) We reverse the sign of one intra-cell hopping in both four-site and eight-site cells
		as highlighted by the red lines.
		(d) Illustration of the tight-binding model on the decagonal quasicrystalline lattice,
		where there are 3010 lattice sites.} 
	\label{FigSM_C10_lattice}
\end{figure*}

We now apply the method previously introduced for constructing an octagonal quasicrystalline structure to
develop a tight-binding model on the Penrose tiling decagonal quasicrystalline lattice.
We first generate the lattice using the cut-and-project method, 
with a cell positioned at each lattice vertex. Each initial vertex contains ten sites, uniformly
distributed on a small circle centered at the vertex, with a radius of $R_c=0.2$ and a polar angle of 
$\theta_m=(2m-1)\pi/10$ for $m=1,\dots,10$ (the quasicrystalline edge length is set to one).
The inter-cell bonds are introduced based on quasicrystalline edges.
The sites that are not linked by any inter-cell bond are removed, with 
the remaining sites within each cell being connected by intro-cell bonds. 
The resulting lattice structure is illustrated in Fig.~\ref{FigSM_C10_lattice}(a).

To enforce chiral symmetry, we modify some cell structures so that all polygons 
enclosed by bonds are even-sided. Specifically, when the intersection angle 
between quasicrystal edges is equal to $3\pi/5$, we replace the two sites within the angular region 
by one site on the angular bisector, which remains on the small circle.
Figure~\ref{FigSM_C10_lattice}(b) plots all types of modified cells.

In addition, we reverse the sign of a hopping in the four-site and 
eight-site cells [see Fig.~\ref{FigSM_C10_lattice}(c)] so as to 
open the bulk energy gap of the tight-binding Hamiltonian 
when the intra-cell hoppings dominate with $t_0/t_1\gg1$.
The resulting decagonal quasicrystalline structure is shown 
in Fig.~\ref{FigSM_C10_lattice}(d).

\begin{figure*}[htp]
	\includegraphics[width=0.9\linewidth]{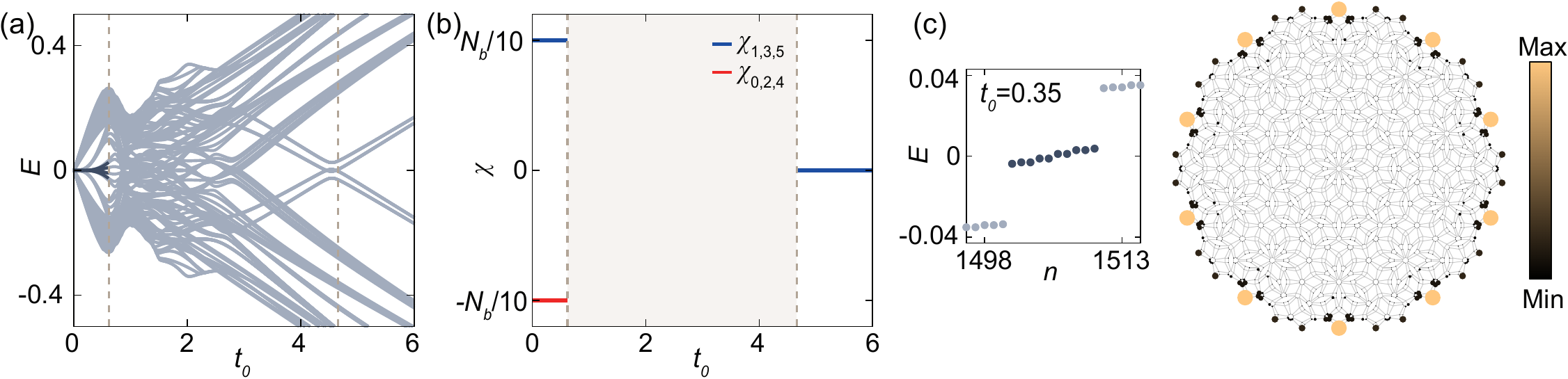}
	\caption{{Results for the tight-binding model on the decagonal quasicrystalline lattice.} 
		(a) Energy spectrum as a function of $t_0$. When $0<t_0<0.62$, 
		ten corner modes (dark blue lines) emerge in the bulk gap. 
		(b) Topological invariants $\chi_p$ versus $t_0$ with $p=0,\dots,5$. 
		All $\chi_p{\text{'s}}$ are nonzero in the regime with corner modes, and $\chi_p=\pm N_b/10$ 
		with $N_b=90$. 
		(c) Energy spectrum versus state index $n$ and spatial distribution of 
		the local DOS at zero energy at $t_0=0.35$.
		Here, we set $t_1=1$, $\lambda=2$, and set the minimal lengths of intra-cell 
		and inter-cell bonds to the value of $0.124$ and $0.620$, respectively.
	} 
	\label{FigSM_C10_E_chi}
\end{figure*}

Similar to the octagonal quasicrystalline case, the Hamiltonian on the decagonal quasicrystalline lattice
respects the time-reversal symmetry, chiral symmetry and the ten-fold rotational symmetry.
The energy spectrum in Fig.~\ref{FigSM_C10_E_chi}(a) demonstrates the existence of ten 
zero-energy modes when $0<t_0<0.62$. These modes are mainly localized at the corners
as revealed by the local DOS at zero energy in Fig.~\ref{FigSM_C10_E_chi}(c). 
However, in contrast to the octagonal quasicrystalline case, no zero-energy modes occur 
in the trivial case, which is consistent with our analysis.

\begin{figure*}[htp]
	\includegraphics[width=0.9\linewidth]{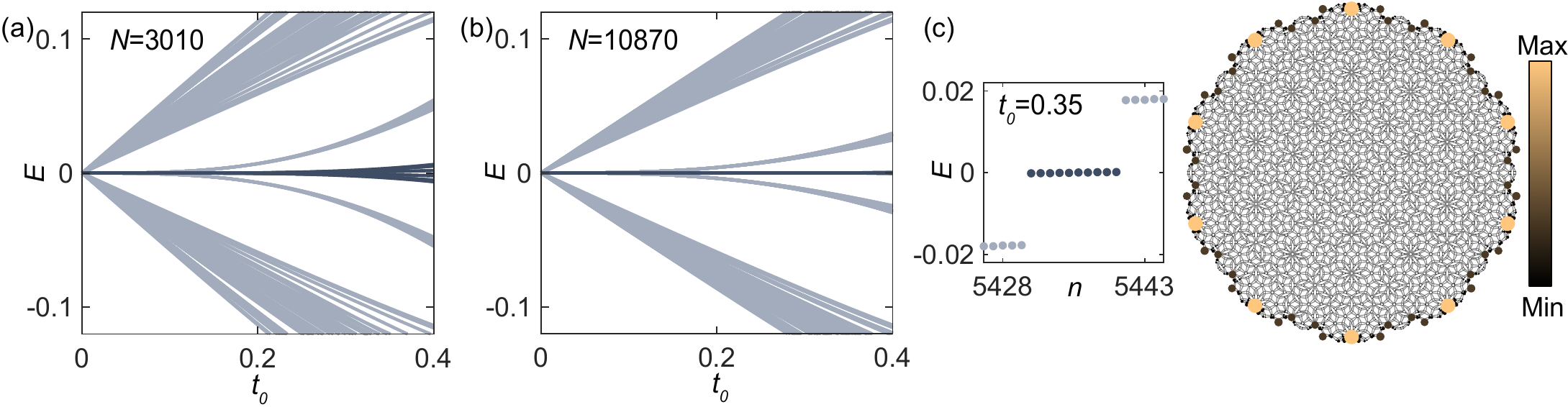}
	\caption{Results for the decagonal quasicrystalline lattice in a larger scale. 
		(a)--(b) Energy spectra of the tight-binding Hamiltonian on a decagonal quasicrystalline lattice 
		as a function of $t_0$ with (a) 3010 and (b) 10870 sites. 
		(a) is a zoomed-in view of Fig.~\ref{FigSM_C10_E_chi}(a) near zero energy. 
		(c) Energy spectrum versus the state index $n$ and spatial distribution of the local DOS at zero energy for $t_0= 0.35$. 
		Here, all system parameters are the same as those in Fig.~\ref{FigSM_C10_E_chi}.
	} 
	\label{Fig_reply_C10_more_site}
\end{figure*}

Similarly, due to rotational symmetry, we define a $\mathbb{Z}$ invariant 
for each eigenspace of the rotational operator $U_{C_{10}}$
as the difference in the number of occupied states below zero energy for the momentum 
space Hamiltonian $H(k)$ between $k=0$ and $k=\pi$. 
Due to time-reversal symmetry, two eigenspaces with eigenvalues $\omega_p$ and $\omega_{10-p}$ 
are related by $\kappa$, leading to $\chi_p=\chi_{10-p}$ with $p=1,\dots,4$.
The conservation of the total number of occupied states results in  
$\sum_{p}\chi_p=0$. Fig.~\ref{FigSM_C10_E_chi}(b) shows that 
the region with corner modes for $0<t_0<0.62$ exhibits nonzero $\chi_p$, which  
equal $\pm N_b/10$ with $N_b$ being the number of boundary sites. 
Moreover, the degeneracy of corner modes in Fig.~\ref{FigSM_C10_E_chi}(c) can be improved via increasing the system size. 
Now, we compute the energy spectrum as a function of $t_0$ for a larger system with 10870 sites. 
Compared to the original case with 3010 sites [see Fig.~\ref{Fig_reply_C10_more_site}(a)], 
the degeneracy of the zero-energy corner modes is significantly improved [see Figs.~\ref{Fig_reply_C10_more_site}(b) and (c)].

%%%%%%%%%%%%%%%%%%%%%%%%%%%%%%%
%\bibliography{reference}
%apsrev4-2.bst 2019-01-14 (MD) hand-edited version of apsrev4-1.bst
%Control: key (0)
%Control: author (8) initials jnrlst
%Control: editor formatted (1) identically to author
%Control: production of article title (0) allowed
%Control: page (0) single
%Control: year (1) truncated
%Control: production of eprint (0) enabled
%

%%%%%%%%%%%%%%%%%%%%%%%%%%%%%%%%%%%
%\hfill \break

%%\newpage
%\onecolumngrid
%
%\begin{center}
%	\large{\textbf{End Matter}}
%\end{center}
%\twocolumngrid
%\appendix
%\renewcommand{\theequation}{A\arabic{equation}}
%\setcounter{equation}{0}

\end{document}